\documentclass[preprint,12pt,authoryear]{elsarticle}

\usepackage{amssymb}
\usepackage{amsmath}
\usepackage{latexsym,epsfig,graphicx,epstopdf,amsmath,amssymb,amscd,undertilde}
\usepackage{multirow,psfrag,paralist,dsfont,url}
\usepackage[titletoc]{appendix}
\usepackage{float}
\usepackage{amsthm}
\usepackage{statmath}
\usepackage{prodint}
\usepackage{natbib}
\usepackage[dvipsnames]{xcolor}
\usepackage{makecell}
\usepackage[american]{babel}
\usepackage[skip=2pt]{caption} % example skip set to 2pt
\usepackage[section]{placeins}
\usepackage{xr}

\usepackage{mathtools}
\usepackage[mathscr]{euscript}
\allowdisplaybreaks
\newcommand{\betavec}{{\boldsymbol{\beta}}}

\newcommand{\lambdavec}{{\boldsymbol{\Lambda}}}

\newcommand{\pr}{{\rm Pr}}

\newcommand{\Var}{{\rm Var}}

\newcommand{\dgoto}{\overset{\text{D}}{\rightarrow}}
\newcommand{\Pgoto}{\overset{\text{P}}{\rightarrow}}
\newcommand{\asgoto}{\overset{\text{a.s.}}{\longrightarrow}}

\newcommand{\xvec}{\boldsymbol{x}}

\newcommand{\loglik}{\mathcal{L}}

\newcommand{\R}{\mathcal{R}}
\newcommand{\A}{\mathcal{A}}

\newcommand{\D}{\mathcal{D}}
\newcommand{\C}{\mathcal{C}}

\newcommand{\bigO}{\mathcal{O}}

\newcommand{\Bern}{\text{Bernoulli}}

\newcommand{\truebeta}{\boldsymbol{\beta}}
\newcommand{\betahat}{\widehat{\boldsymbol{\beta}}}
\newcommand{\betahatB}{\widehat{\boldsymbol{\beta}}_{\text{b}}}
\newcommand{\betahatE}{\widehat{\boldsymbol{\beta}}_{\text{e}}}
\newcommand{\betahatPB}{\widehat{\boldsymbol{\beta}}_{\text{pb}}}

\newcommand{\sbetahat}{\widehat{{\beta}}}
\newcommand{\sbetahatB}{\widehat{{\beta}}_{\text{b}}}
\newcommand{\sbetahatE}{\widehat{{\beta}}_{\text{e}}}
\newcommand{\sbetahatPB}{\widehat{{\beta}}_{\text{pb}}}

\newcommand{\lambdahat}{\widehat{\lambda}}
\newcommand{\Lambdahat}{\widehat{\Lambda}}
\newcommand{\lambdavechat}{\widehat{\lambdavec}}
\newcommand{\ivec}{{\boldsymbol{i}}}

\newtheorem{prethm}{Theorem}
\newenvironment{thm}%
  {\begin{prethm}\upshape}{\end{prethm}}

\newtheorem{prelemma}{Lemma}
{\begin{prelemma}\upshape}{\end{prelemma}}

\newtheorem{prerem}{Remark}
\newenvironment{remark}%
{\begin{prerem}\upshape}{\end{prerem}}

\newtheorem{preassump}{Assumption}
\newenvironment{assumption}%
{\begin{preassump}\upshape}{\end{preassump}}

\mathtoolsset{showonlyrefs=true}

\journal{Computational Statistics $\&$ Data Analysis}

\begin{document}

\begin{frontmatter}



\title{An Accurate Computational Approach for Partial Likelihood Using Poisson-Binomial Distributions} %% Article title


\author[vt]{
Youngjin Cho} %% Author name
%\ead{youngjin@vt.edu}
%\ead{youngjin@vt.edu}
\author[vt]{
Yili Hong} %% Author name
%\ead{yilihong@vt.edu}
%\ead{yilihong@vt.edu}
\author[vt]{Pang Du\corref{cor}} %% Author name
\ead{pangdu@vt.edu}

\cortext[cor]{Corresponding author}
\fntext[general]{The proposed method is implemented in an R package \texttt{ExactCoxPBD} available at \url{https://github.com/Stat-Y/ExactCoxPBD}.}
%\address{Department of Statistics, Virginia Tech, 250 Drillfield\\
%Drive, Blacksburg, 24061, Virginia, United States}
\affiliation[vt]{organization={Department of Statistics, Virginia Tech},%Department and Organization
            addressline={250 Drillfield Drive}, 
            city={Blacksburg},
            postcode={24061}, 
            state={Virginia},
            country={United States}}

%% Abstract
\begin{abstract}
In a Cox model, the partial likelihood, as the product of a series of conditional probabilities, is used to estimate the regression coefficients. In practice, those conditional probabilities are approximated by risk score ratios based on a continuous time model, and thus result in parameter estimates from only an approximate partial likelihood. Through a revisit to the original partial likelihood idea, an accurate partial likelihood computing method for the Cox model is proposed, which calculates the exact conditional probability using the Poisson-binomial distribution. New estimating and inference procedures are developed, and theoretical results are established for the proposed computational procedure. Although ties are common in real studies, current theories for the Cox model mostly do not consider cases for tied data. In contrast, the new approach includes the theory for grouped data, which allows ties, and also includes the theory for continuous data without ties, providing a unified framework for computing partial likelihood for data with or without ties. Numerical results show that the proposed method outperforms current methods in reducing bias and mean squared error, while achieving improved confidence interval coverage rates, especially when there are many ties or when the variability in risk scores is large. Comparisons between methods in real applications have been made. 
\end{abstract}


\iffalse
\begin{highlights}
\item Proposed a procedure to optimize the exact partial likelihood in the Cox model.
\item Efficiently computed Poisson-binomial distribution in the exact partial likelihood.
\item Proved consistency and asymptotic normality for both with/without ties scenarios.
\item Compared to other methods, the proposed method has notably less bias.
\item The proposed method excels others in data with many ties or high covariate variation.
\end{highlights}
\fi

%% Keywords
\begin{keyword}
Cox Model \sep Proportional Hazards Model \sep Breslow Estimator \sep Efron Estimator \sep Kalbfleisch-Prentice Correction %\sep Martingale Central Limit Theorem 



\end{keyword}

\end{frontmatter}



\section{Introduction}
\label{sec:intro}
Ever since its birth half a century ago, the Cox model \citep{Cox1972} has been arguably the most widely used method to analyze survival data with covariates. Since the literature for Cox models is too vast to give a comprehensive review here, we only name a few widely-used monographs here, such as \citet{FlemingHarrington1991}, \citet{TherneauGrambsch2000},
\citet{KalbfleischPrentice2002}, and \citet{klein03}.

Consider a typical survival data setting where observations are denoted by $\{t_i, \delta_i, \xvec_i\}$, $i=1, \dots, n$. Here, $n$ is the number of subjects in the study,  $t_i$ is the observed event time or censoring time, and $\delta_i$ is the event indicator that equals to 1 if the observed time is from an event and 0 if the event time is censored. The covariate vector is denoted by $\xvec_i=(x_{i1},\dots,x_{id})^{\text{T}}$, where $d$ is the number of covariates. The Cox model \citep{Cox1975} incorporates the covariate effects into the hazard function as
$
\lambda(t; \xvec_i)=\lambda_0(t)\exp(\xvec_i^{\text{T}}\truebeta),
$
where $\lambda_0(t)$ is the unknown baseline hazard function and $\truebeta=(\beta_{1},\dots,\beta_{d})^{\text{T}}$ is the vector of regression coefficients. Let $\Lambda_0(t)=\int_0^t \lambda_0(s) ds$ be the baseline cumulative hazard function, which is often estimated as a step function. Denote the cumulative hazard function for a specific $\xvec_i$ by $\Lambda(t; \xvec_i)=\int_0^t \lambda(s; \xvec_i) ds$. The regression coefficients $\truebeta$ are typically estimated by the partial likelihood.

To revisit the idea of partial likelihood, let $t_{(j)}$, $j=1,\dots, k$, be the $k$ ordered {\it distinct} events times in data $\{t_i, \delta_i, \xvec_i\}_{i=1}^n$. 
To clarify, when referring to something up to $t$, this includes all times from $0$ to $t^-$ but explicitly excludes $t$ itself.  Let $\R(t_{(j)})$ be the at-risk set containing all the subjects that survived up to time $t_{(j)}$. Let  $n_j\equiv|\R(t_{(j)})|$. When there are no ties, \citet{Cox1975} constructs the partial likelihood as $L(\truebeta)=\prod_{j=1}^k L_j(\truebeta)$,
where
\begin{align}\nonumber
L_j(\truebeta)&=\frac{\pr( \text{unit }j_1\text{ failed at }t_{(j)} \big| \,n_j\text{ units survived up to }t_{(j)})}{\pr ( 1 \text{ out of } n_j\text{ units failed at }t_{(j)} \big|\, n_j\text{ units survived up to }t_{(j)})}\\\label{eqn:partial_j.original}
&=\frac{d\Lambda(t_{(j)}; \xvec_{j_1}) \prod_{i\in \R(t_{(j)})\setminus \{j_1\} }(1-d\Lambda(t_{(j)}; \xvec_{i}))}{\sum_{i\in \R(t_{(j)})} \left\{d\Lambda(t_{(j)}; \xvec_{i}) \prod_{l\in \R(t_{(j)})\setminus \{i\} }(1-d\Lambda(t_{(j)}; \xvec_{l}))\right\}}\,\,
\end{align}
is the partial likelihood contribution from observations at time $t_{(j)}$, and $j_1$ is the index for the only failed subject at time $t_{(j)}$. The expression \eqref{eqn:partial_j.original} from \citet{Cox1975} represents the accurate likelihood contribution from the observation at time $t_{(j)}$. We refer to \eqref{eqn:partial_j.original}  as the {\it accurate partial likelihood} (APL). Despite its accuracy, \eqref{eqn:partial_j.original} is not widely used in practice due to its computational difficulty in the denominator.

Ignoring higher order terms under a continuous failure time model when no ties are present, \eqref{eqn:partial_j.original} can be approximated by 
\begin{align}\label{eqn:PL.CA.j}
L_j(\truebeta)=\frac{\exp(\xvec_{j_1}^{\text{T}}\truebeta)}{\sum_{i\in \R(t_{(j)})}\exp(\xvec_i^{\text{T}}\truebeta)};
\end{align}
see, e.g., Page 140 of \citet{FlemingHarrington1991}.  
For decades, \eqref{eqn:PL.CA.j} has been used in practice and implemented in major software packages. In many textbooks and research papers, \eqref{eqn:PL.CA.j} is often directly introduced as the partial likelihood and the original APL idea gets buried. The goal of this paper is to re-visit the idea of APL with some new development from another area, by realizing that the denominator of \eqref{eqn:partial_j.original} is in the form of the probability mass function of a Poisson-binomial (PB) distribution. The PB distribution describes the sum of independent but non-identically distributed random indicators \citep{Hong2013,PBD_2018}.

When ties are present, the calculation of the APL gets even harder and more time consuming. To incorporate ties into Cox models, several approximating approaches have been proposed. In essence, they are all direct extensions of \eqref{eqn:PL.CA.j} to the with-tie scenario, with some additional ad hoc approximations to include the contributions from all the event times tied at a time point. Suppose $d_j$ is the number of failures at time $t_{(j)}$. The Cox correction \citep{Cox1972} averages the corresponding partial likelihoods over all the size-$d_j$ subsets of the at risk set at time $t_{(j)}$. The Kalbfleisch-Prentice correction \citep{KalbfleischPrentice1973} averages on all the possible permutations of the $d_j$ underlying event times. The Breslow correction \citep{Breslow1974} is a simplified version of the Kalbfleisch-Prentice correction assuming an equal contribution from each permutation. The Efron correction \citep{Efron1977} can be viewed as a centerized version of the Breslow correction. \citet{TherneauGrambsch2000} provided an overview of tie corrections in partial likelihood and also introduced various Cox regression based models. \citet{HertzRockhill1997} and \citet{ScheikeSun2007} used numerical studies to compare performances of different tie correction methods, and similar simulation settings are adopted in this paper. These methods are now standards for tie corrections in Cox models and well-implemented in all the major statistical software. 

However, they may not be the ideal approaches for tie corrections. Firstly, they are all based on \eqref{eqn:PL.CA.j}, which is already an approximation to the APL and derived actually for continuous failure times without ties. Secondly, the various averaging approaches are really crude ways to count the contributions of all the tied event times without fully incorporating the distributional differences among these ties. This creates further deviations from the APL. Furthermore, as far as we know, there are no formal investigations on the asymptotic properties of these correction methods under consideration for how large the permissible ties can be in the model.

In this paper, we propose a new computationally efficient method to calculate the APL based on the PB distribution development in \citet{Hong2013}. The key idea is that the denominator of the APL is exactly in the form of the probability mass function of a PB distribution, regardless the presence of ties or not. We use the method in \citet{Hong2013} to compute the PB probability mass function, which is based on the discrete Fourier transformation of the characteristic function. Alternatively, one can also use the convolution-based method in \citet{PBD_2018} based on the direct convolution or the divide-and-conquer fast Fourier transform tree convolution. As far as we know, the idea of a direct and exact computation of the APL for the Cox model in this paper is completely new to the literature.

We consider the common scenario where ties are caused by rounding or grouping of underlying continuous failure times, which we refer to as the grouped continuous failure time model. Our first result shows that all the aforementioned common methods, namely, the expression \eqref{eqn:PL.CA.j}, the Cox correction, the Kalbfleisch-Prentice correction, the Breslow correction, and the Efron correction, are all approximations to the PB distribution approach based on Poisson approximation or approximation by enumeration averaging. Under the grouped continuous failure time model, we first establish the consistency and asymptotic normality for the Breslow estimator, which can be easily extended to the estimators from the other existing correction methods. For our PB distribution approach, we establish the consistency and asymptotic normality for its estimator under both the grouped continuous failure time model (ties present) and the ungrouped one (no ties). Note that although \cite{Prentice2003MixedDA} provided theoretical results for the Breslow estimator in the presence of ties, they did not address how large the order of ties can be for the model, whereas we provide practical insights into the permissible order of ties. In simulations, we compare the performance of our PB distribution approach with those of the existing approaches such as the Breslow correction and the Efron correction. Our result shows that our  approach has much lower biases and mean squared errors, as well as higher confidence interval coverage rates, than the existing methods for survival data with heavy ties or high-variation covariates. Our real data examples further confirm our findings in the theory and simulation.

In summary, our new PB distribution based approach to compute the partial likelihood accurately has the following distinct contributions: (1) we show that all the existing tie correction methods are approximations to our exact approach; (2) we derive the consistency and asymptotic normality of these methods under this unified framework, suggesting allowable order of ties, which has been lacking in the literature despite their popularity and long history; (3) we show that the proposed approach possesses the same asymptotic properties and demonstrate its clear numerical advantage in terms of reduced bias and mean squared error, along with enhanced confidence interval coverages.

The rest of the paper is organized as follows. In Section~\ref{sec:cox.model}, we introduce the Cox model and the PB distribution. In Section~\ref{sec:partial_like}, we obtain the accurate partial likelihood for the Cox model using the PB distribution, and propose a new method to estimate the coefficients and the baseline hazard function for the Cox model. In Section \ref{sec:stat_theory}, we develop statistical theory for the existing methods and our new method. In Section \ref{sec:simul}, we evaluate the performance of our method in simulated data and  compare its performance with others. In Section \ref{sec:data_analysis}, we analyze survival datasets with our new method and the existing ones.
Lastly, in Section~\ref{sec:Conclusion}, we conclude the paper with recommendations and some areas for future research. Technical proofs and R code for numerical results are collected in the online supplementary materials.

%%%%%%%%%%%%%%%%%%%%%%%%%%%%%%%%%%%%%%%%%%%%%%%%%%%%%%%%%%%%%%%%%%%%%%%%%%%%%%%%%
\section{The Cox Model}\label{sec:cox.model}
%%%%%%%%%%%%%%%%%%%%%%%%%%%%%%%%%%%%%%%%%%%%%%%%%%%%%%%%%%%%%%%%%%%%%%%%%%%%%%%%%
%%%%%%%%%%%%%%%%%%%%%%%%%%%%%%%%%%%%%%%%%%%%%%%%%%%%%%%%%%%%%%%%%%%%%%%%%%%%%%%%%
\subsection{The Underlying Continuous Failure Time Model}\label{sec:underlying_model}
%%%%%%%%%%%%%%%%%%%%%%%%%%%%%%%%%%%%%%%%%%%%%%%%%%%%%%%%%%%%%%%%%%%%%%%%%%%%%%%%%

In a continuous failure time model, the underlying event times $\tilde{T}_i \in (0, \infty)$, $i=1,\dots, n$, are from a continuous distribution. Here, $\tilde{T}_1,\dots, \tilde{T}_n$ are assumed to be conditionally independent given the covariates $\{\xvec_i\}_{i=1}^n$. Recall the general form of the Cox model: for $i=1,\dots,n$,
\begin{align}\label{eqn:conti_cox}
d\Lambda(t; \xvec_i)= \pr(\tilde{T}_i \in [t, t+dt) | \xvec_i, \tilde{T}_i \geq t)=\exp(\xvec_i^{\text{T}}\truebeta)d\Lambda_0(t).
\end{align}
The cumulative hazard and survival functions are respectively $\Lambda(t;\xvec_i)=\int_0^td\Lambda(s;\xvec_i)=\exp(\xvec_i^{\text{T}}\truebeta)\Lambda_0(t)$ and $
S(t;\xvec_i)= \pr(\tilde{T}_i > t|\xvec_i)=\exp(-\Lambda(t;\xvec_i))$. Under the continuous failure time model, the baseline hazard function can be written as
$d\Lambda_0(t)=\lambda_0(t)dt$.
Let $\zeta \in (0, \infty)$ be the ending time of the study and $C_i \in (0, \zeta],i=1,\dots,n$, be non-informative right censoring times, which are conditionally independent from each other and from $\{\tilde T_i\}_{i=1}^n$ given the covariates $\{\xvec_i\}_{i=1}^n$. Assume that the covariates are deterministic, allowing the conditioning argument for $\{\xvec_i\}_{i=1}^n$ to be omitted for simplicity.

%%%%%%%%%%%%%%%%%%%%%%%%%%%%%%%%%%%%%%%%%%%%%%%%%%%%%%%%%%%%%%%%%%%%%%%%%%%%%%%%%
\subsection{The Grouped Continuous Failure Time Model}\label{sec:CMG}
%%%%%%%%%%%%%%%%%%%%%%%%%%%%%%%%%%%%%%%%%%%%%%%%%%%%%%%%%%%%%%%%%%%%%%%%%%%%%%%%%
In some scenarios, it is possible that the event times are rounded or grouped, leading to ties in the event times. Let $\lceil a \rceil$ be the smallest integer which is not smaller than $a$. Let $\tau \in (0, \infty)$ be a grouping parameter such that event times and censoring times are discretized as
\begin{equation}\label{eqn:grouping.model.discretize}
\tilde{T}_i^{\ast} = \tau \lceil \tilde{T}_i / \tau \rceil \in \Omega_{\text{G}}=\{\tau, 2\tau, \dots\} \text{ and } C_i^{\ast}=\tau \lceil C_i / \tau \rceil\in \Omega=\{\tau, 2\tau, \dots, {\color{black}\zeta}\},
\end{equation}
where we assume, without loss of generality, that the ending time $\zeta$ is a multiple of $\tau$. The observed time is $T_i^{\ast}=\mbox{min}(\tilde{T}^{\ast}_i, C_i^{\ast}) \in \Omega$, where $\delta_i=\mathds{1}(\tilde{T}_i^{\ast} \leq C_i^{\ast})$ is the indicator of an event. That is, when $\delta_i=1$, the $i$th unit had the event and when $\delta_i=0$, the $i$th unit was censored. As $\Omega_{\text{G}}$ is the support for grouped times, $\Omega$ is the support for the observed times, which is a finite subset of $\Omega_{\text{G}}$. In this paper, we assume ties are generated by the grouping of an underlying continuous random variable.
From the discretized distribution, for $t \in \Omega_{\text{G}}$, we have
\begin{align}\label{eqn:deriving.grouping.hf.1}
  \pr(\tilde{T}^{\ast}_i \geq t | \xvec_i)&=\pr(\tilde{T}_i > t-\tau | \xvec_i)=S(t-\tau;\xvec_i),
\\
\label{eqn:deriving.grouping.hf.2}
\pr(\tilde{T}^{\ast}_i \in [t, t+dt) | \xvec_i) & =\pr(\tilde{T}_i \in (t-\tau, t] | \xvec_i)=S(t-\tau;\xvec_i)-S(t;\xvec_i),
\end{align}
for $i=1,\dots,n$. From \eqref{eqn:deriving.grouping.hf.1} and \eqref{eqn:deriving.grouping.hf.2}, the hazard function of $\tilde{T}_i^{\ast}$ is \begin{align}\label{eqn:discretized_cox}
d\Lambda^{\ast}(t;\xvec_i)&=\pr(\tilde{T}^{\ast}_i \in [t, t+dt) | \xvec_i,\tilde{T}^{\ast}_i \geq t)=\frac{\pr(\tilde{T}^{\ast}_i \in [t, t+dt) | \xvec_i)}{\pr(\tilde{T}^{\ast}_i \geq t | \xvec_i)}\\\nonumber
&=1-\exp\left(-\exp( \xvec_i^{\text{T}} \truebeta ) d\Lambda_0^{\ast}(t)\right), \quad i=1,\dots,n,
\end{align} where the baseline hazard function is $d\Lambda_0^{\ast}(t)=\Lambda_0(t)-\Lambda_0(t-\tau)$ for $t \in \Omega_{\text{G}}$, $d\Lambda_0^{\ast}(t)=0$ for $t \notin \Omega_{\text{G}}$, and the baseline cumulative hazard function is \begin{align}\label{eqn:discretized_cox_base_chf}
%d\Lambda_0^{\ast}(t)&=d\Lambda_0(t) \times \Indfun(t\in \{t_1,\dots,t_k\})\\
\Lambda_0^{\ast}(t)&= \int_{0}^{t}d\Lambda_0^{\ast}(s) =\sum_{s\leq t, s \in \Omega_{\text{G}}}d\Lambda_0^{\ast}(s).
\end{align}
As $\Lambda_0(0)=0$, we have $\Lambda_0^{\ast}(t)=\Lambda_0(t)$ for $t \in \Omega_{\text{G}} \cup \{0\}$. The cumulative hazard function and survival function of $\tilde{T}^{\ast}_i$ are respectively
$
\Lambda^\ast(t;\xvec_i)=\int_0^td\Lambda^\ast(s;\xvec_i)=\sum_{s\leq t, s \in \Omega_{\text{G}}}d\Lambda^\ast(s;\xvec_i)$
and
$S^{\ast}(t;\xvec_i) =  \pr(\tilde{T}^{\ast}_i > t|\xvec_i) =$\\ $\prod_{s \leq t, s \in \Omega_{\text{G}}}(1-d\Lambda^{\ast}(s;\xvec_i)) =  \exp(-\exp(\xvec_i^{\text{T}}\truebeta)$ $\Lambda_0^\ast(t))$. One can see that $S^{\ast}(t;\xvec_i)$ $=S(t;\xvec_i)$ for $t \in \Omega_{\text{G}} \cup \{0\}$.

%%%%%%%%%%%%%%%%%%%%%%%%%%%%%%%%%%%%%%%%%%%%%%%%%%%%%%%%%%%%%%%%%%%%%%%%%%%%%%%%%
\subsection{The Fitted Model}\label{sec:Discretized}
%%%%%%%%%%%%%%%%%%%%%%%%%%%%%%%%%%%%%%%%%%%%%%%%%%%%%%%%%%%%%%%%%%%%%%%%%%%%%%%%%
In a typical semiparametric inference setting, the baseline cumulative hazard function is estimated as a step function, regardless of whether data are with grouping or not. Hence, the fitted model is essentially a discretized version of the continuous model in \eqref{eqn:conti_cox}.

Let $t_{(j)}$, $j=1,\dots, k$, be the ordered distinct events times as defined in Section \ref{sec:intro} and let $\Omega^{\mathcal{K}}=\{t: t=t_{(j)}, j=1,\dots,k\} \subseteq (0, \infty)$ be the set of distinct event times. It is worth reiterating no matter how data are generated (i.e., from a continuous model with or without grouping), the fitted Cox model is always a discretized model, because the estimated cumulative hazard function is a step function which has jumps at those time points in $\Omega^{\mathcal{K}}$. Under the grouped continuous failure time model, we have $\Omega^{\mathcal{K}} \subseteq \Omega \subseteq \Omega_{\text G} \subseteq (0, \infty)$.

%%%%%%%%%%%%%%%%%%%%%%%%%%%%%%%%%%%%%%%%%%%%%%%%%%%%%%%%%%%%%%%%%%%%%%%%%%%%%%%%%
\subsection{Poisson-Binomial Distribution}\label{sec:pbd}
%%%%%%%%%%%%%%%%%%%%%%%%%%%%%%%%%%%%%%%%%%%%%%%%%%%%%%%%%%%%%%%%%%%%%%%%%%%%%%%%%
Recall that $n_j$ is the number of subjects in the at-risk set $\R(t_{(j)})$ at time $t_{(j)}$, and $d_j$ is the number of subjects in the event set $\D(t_{(j)})$ at time $t_{(j)}$. Here $d_j$ can be larger than one so that ties are possible. For each $i\in \R(t_{(j)})$, given the history up to time $t_{(j)}$, the probability $p_{ij}$ for subject $i$ to have an event at time $t_{(j)}$ based on the fitted model is defined as
$p_{ij}=p_{ij}(\truebeta, \lambda_j)\equiv d\Lambda^{\ast}(t_{(j)};\xvec_i)$
by \eqref{eqn:discretized_cox}, where $\lambda_j$ is defined as $\lambda_j \equiv d\Lambda_0^{\ast}(t_{(j)})$ by \eqref{eqn:discretized_cox_base_chf}. From now until the end of this section, all distributional and independence derivations are conditional on the history up to $t_{(j)}$, but for simplicity, we omit this conditioning and use abused notation. We can define a random indicator $I_{ij} \sim \Bern(p_{ij})$, where $\Bern(p_{ij})$ indicates a Bernoulli distribution with parameter $p_{ij}$. In this case, the number of events $I_j$ at time $t_{(j)}$ can be written as
$I_{j}=\sum_{i\in \R(t_{(j)})} I_{ij}$.
Because $p_{ij}$’s are different, the distribution of $I_j$ is not necessarily a binomial distribution. In general, $I_{j}$ follows a PB distribution. The probability mass function of a PB distribution can be computed using enumeration, the method of discrete Fourier transform of characteristic function, an approximation method such as the Poisson approximation \citep{Hong2013}, or the convolution-based method \citep{PBD_2018}.

The enumeration method computes the probability mass function of $I_j$ as follows,
\begin{align}\label{eqn:PB.enum}
\pr(I_j=d_j)=\sum_{\A_{d_j}\in\mathscr{F}_{d_j}}\left\{\prod_{i\in \A_{d_j}}{p_{ij}}\right\}\left\{\prod_{i\in \R(t_{(j)})\setminus \A_{d_j}}(1-{p_{ij}})\right\},
\end{align}
where $\mathscr{F}_{d_j}$ is the set of all subsets of $d_j$ individuals that can be selected from $\R(t_{(j)})$. That is, each set $\A_{d_j} \in \mathscr{F}_{d_j}$ has $d_j$ number of elements and $|\mathscr{F}_{d_j}|=\binom{n_j}{d_j}$. The method of discrete transform of characteristic function computes the probability mass function of $I_j$ as
$\pr(I_j=d_j)=(n_j+1)^{-1}\sum_{l=0}^{n_j}\exp(-\ivec\omega_j ld_j)z_{lj}$,
where $\omega_j=2\pi/(n_j+1)$, $\ivec=\sqrt{-1}$, and $z_{lj}=\prod_{i \in \R(t_{(j)})}\left\{1-p_{ij}+p_{ij}\exp(\ivec\omega_j l)\right\}.$ The method can be done efficiently using the fast Fourier transform, and it is available in the R package ``$\texttt{poibin}$" by \citet{Hong2013poibin}. The Poisson approximation to the probability mass function of a PB distribution is,
\begin{align}\label{eqn:pbd.pa}
\pr(I_j=d_j)\approx\frac{\mu_j^{d_j} \exp(-\mu_j)}{d_j!},
\end{align}
where $\mu_j=\sum_{i\in \R(t_{(j)})}p_{ij}$ is the mean of $I_j$. 

%%%%%%%%%%%%%%%%%%%%%%%%%%%%%%%%%%%%%%%%%%%%%%%%%%%%%%%%%%%%%%%%%%%%%%%%%%%%%%%%%
\section{Partial Likelihood and Parameter Estimation}\label{sec:partial_like}
%%%%%%%%%%%%%%%%%%%%%%%%%%%%%%%%%%%%%%%%%%%%%%%%%%%%%%%%%%%%%%%%%%%%%%%%%%%%%%%%%

%%%%%%%%%%%%%%%%%%%%%%%%%%%%%%%%%%%%%%%%%%%%%%%%%%%%%%%%%%%%%%%%%%%%%%%%%%%%%%%%%
\subsection{The Original Idea of Partial Likelihood}
%%%%%%%%%%%%%%%%%%%%%%%%%%%%%%%%%%%%%%%%%%%%%%%%%%%%%%%%%%%%%%%%%%%%%%%%%%%%%%%%%
Consider the general situation where ties could be present or not. The APL function is
$L(\truebeta, \lambdavec)=\prod_{j=1}^k L_j(\truebeta, \lambda_j)=\prod_{j=1}^k A_j(\truebeta, \lambda_j)/B_j(\truebeta, \lambda_j)$,
where $\lambdavec=(\lambda_1,\dots,\lambda_k)^{\text{T}}$ and
\begin{align}\label{eqn:partial_j}
L_j(\truebeta, \lambda_j)&=\frac{\pr( \text{units }j_1,\dots,j_{d_j}\text{ had event at }t_{(j)} \big| \,n_j\text{ units survived up to }t_{(j)})}{\pr ( d_j\text{ out of } n_j\text{ units had event at }t_{(j)} \big|\, n_j\text{ units survived up to }t_{(j)})}.
\end{align} Here $j_1,\dots,j_{d_j}$ are the $d_j$ individuals in $\D(t_{(j)})$. The accurate calculation for $A_j(\truebeta, \lambda_j)$ is
\begin{align}\label{eqn:Nj}
A_j(\truebeta,\lambda_j)=\left\{\prod_{i\in \D(t_{(j)})} p_{ij}(\truebeta,\lambda_j)\right\} \left\{ \prod_{i\in \R(t_{(j)})\setminus \D(t_{(j)}) }\left(1-p_{ij}(\truebeta,\lambda_j)\right)\right\}.
\end{align}
The accurate calculation for $B_j(\truebeta, \lambda_j)$ is
\begin{align}\label{eqn:Dj}
B_j(\truebeta,\lambda_j)=\pr(I_j=d_j),
\end{align}
which is the probability mass function of the PB distribution. 

%%%%%%%%%%%%%%%%%%%%%%%%%%%%%%%%%%%%%%%%%%%%%%%%%%%%%%%%%%%%%%%%%%%%%%%%%%%%%%%%%
\subsection{An Estimation Procedure Based on PB distributions}\label{sec:new_method}
%%%%%%%%%%%%%%%%%%%%%%%%%%%%%%%%%%%%%%%%%%%%%%%%%%%%%%%%%%%%%%%%%%%%%%%%%%%%%%%%%
When there are no ties ($d_j=1$), the expression \eqref{eqn:PL.CA.j} is used to approximate the APL \citep{Cox1972}. In the presence of ties, various corrections can be applied, including the Cox correction \citep{Cox1972}, the Kalbfleisch-Prentice correction \citep{KalbfleischPrentice1973}, the Breslow correction \citep{Breslow1974}, and the Efron correction \citep{Efron1977}. More details about these methods are in \ref{sec:exisingmethods}. It is important to note that all of these existing methods rely on some approximations to both $A_j(\truebeta, \lambda_j)$ and $B_j(\truebeta, \lambda_j)$.

Once the probability mass function of the PB distribution is calculated, we can compute the accurate probability terms \eqref{eqn:partial_j} in the partial likelihood function. However, the probability $p_{ij}$ depends on $\lambdavec$. Let $\lambdavechat=(\lambdahat_{1},\dots,\lambdahat_{k})^{\text{T}}$ be any estimate of $\lambdavec$, which satisfies a mild condition in Assumption \ref{assump:efron}. We use $\lambdavechat$ to substitute $\lambdavec$. Our proposed partial likelihood to be calculated based on PB distribution is
\begin{align}\label{eqn:PL.PB}
L(\truebeta, \lambdavechat)=\prod_{j=1}^{k}\frac{A_j(\truebeta, \lambdahat_{j})}{B_j(\truebeta, \lambdahat_{j})},
\end{align}
where $A_j$ and $B_j$ are given in \eqref{eqn:Nj} and \eqref{eqn:Dj}, respectively. Then the PB distribution estimate of $\truebeta$ is
\begin{align}\label{eqn:PB_betahat}
\betahatPB=\argmax_{\tilde\truebeta} L(\tilde\truebeta, \lambdavechat),
\end{align}
which can be computed optimization algorithms such as the BFGS algorithm (e.g., \citealp{BFGS_book}). The new method uses the exact $A_j(\tilde\truebeta, \lambdahat_{j})$ and $B_j(\tilde\truebeta, \lambdahat_{j})$ for a given $\tilde\truebeta$. Hence, $\betahatPB$ is the maximizer of a more accurate partial likelihood.

The optimization of \eqref{eqn:PB_betahat} requires initial estimates $\lambdavechat$ and $\betahat$ whose choices are flexible. For example, one can use the Efron baseline hazard function estimate in \eqref{eqn:efron}, $\lambdavechat_{\text{e}}=(\lambdahat_{\text{e}1},\dots,\lambdahat_{\text{e}k})^{\text{T}}$, the Breslow baseline hazard function estimate in \eqref{eqn:breslow}, $\lambdavechat_{\text{b}}=(\lambdahat_{\text{b}1},\dots,\lambdahat_{\text{b}k})^{\text{T}}$, or even the Nelson-Aalen baseline hazard function estimate, $\lambdavechat_{\text{na}}=(\lambdahat_{\text{na},1},\dots,\lambdahat_{\text{na},k})^{\text{T}}$ as $\lambdavechat$, where $\lambdahat_{\text{na},j}=d_j/n_j$. Our numerical example in Supplementary Section S3.2 suggests choosing $\lambdavechat_{\text{e}}$ yields a smaller bias comparing to the other methods. So we use $\lambdavechat_{\text{e}}$ as $\lambdavechat$ in our numerical examples in Section \ref{sec:simul} and \ref{sec:data_analysis}. For $\betahat$, one can use Efron estimator $\betahatE$ in \eqref{eqn:efron} or Breslow estimator $\betahatB$ in \eqref{eqn:breslow}. We use $\betahatE$ in our numerical examples in Sections \ref{sec:simul} and \ref{sec:data_analysis}.

Next, we estimate the baseline hazard function. The likelihood for the baseline hazard function is $\prod_{j=1}^k A_j(\truebeta, \lambda_j)$ as shown in page 115 of \citet{KalbfleischPrentice2002}. So, by the BFGS algorithm with the initial values $\lambdahat_{j}$, the baseline hazard function can be updated as $\lambdahat_{\text{pb},j}=\argmax_{\tilde\lambda_j} A_j(\betahatPB,\tilde\lambda_j)$, $j=1,\dots,k$, where the fitted cumulative hazard function is $\widehat{\Lambda}_{\text{pb}}(t) = $\\ $\sum_{j=1}^k \lambdahat_{\text{pb},j} \mathds{1}(t_{(j)} \leq t)$.

%%%%%%%%%%%%%%%%%%%%%%%%%%%%%%%%%%%%%%%%%%%%%%%%%%%%%%%%%%%%%%%%%%%%%%%%%%%%%%%%%
\section{Statistical Properties}\label{sec:stat_theory}
%%%%%%%%%%%%%%%%%%%%%%%%%%%%%%%%%%%%%%%%%%%%%%%%%%%%%%%%%%%%%%%%%%%%%%%%%%%%%%%%%

\subsection{Connections to PB Distribution}\label{sec:connect_pbd}
We first draw some connections between the PB distribution probability and existing partial likelihood approximation methods. Interestingly, all the existing methods are connected to the PB distribution approach, and our results shed theoretical insights on when the existing computing methods tend to work well.

\begin{thm}\label{thm:relation.to.pb}
(\romannumeral 1) the approximate partial likelihood \eqref{eqn:PL.no.ties} is based on the Poisson approximation of $\pr(I_j=1)$ to $B_j(\truebeta,\lambda_j)$, (\romannumeral 2) the Breslow correction in \eqref{eqn:breslow} based on the Poisson approximation of $\pr(I_j=d_j)$ to $B_j(\truebeta,\lambda_j)$, (\romannumeral 3) the Cox correction in \eqref{eqn:exact_Cox} is based on the enumeration in \eqref{eqn:PB.enum} to compute the $B_j(\truebeta,\lambda_j)$, and (\romannumeral 4) the Kalbfleisch-Prentice correction in \eqref{eqn:exact_KP} is based on the enumeration to compute $A_j(\truebeta,\lambda_j)/B_j(\truebeta,\lambda_j)$.
\end{thm}

The proof of Theorem \ref{thm:relation.to.pb} is in Supplementary Section S1.1. As a note, the Efron correction in \eqref{eqn:efron} is of a similar form to the Breslow correction but with some further adjustments. Thus, the Efron correction can be viewed as a more refined Poisson approximation to the PB distribution probability than the Breslow correction. We provide some insights into the accuracy of their Poisson approximations in the following remark.

\begin{remark}\label{rmk:Lecam}
The error bound of the Poisson approximation to the PB distribution can be obtained by the Le Cam theorem \citep{LeCam1960}. The average error is bounded as
\begin{align*}
\frac{1}{n_j}\sum_{d_j=0}^{n_j}\left|\pr(I_j=d_j)-\frac{\mu_j^{d_j} \exp(-\mu_j)}{d_j!}\right|&\leq\frac{2}{n_j}\sum_{i\in \R(t_{(j)})}p_{ij}^2\\
&=\frac{2\sum_{i\in \R(t_{(j)})}(p_{ij}-\bar{p}_{j})^2}{n_j}+2\bar{p}_{j}^2.
\end{align*}
Here $\bar{p}_{j}=\sum_{i\in \R(t_{(j)})}p_{ij}/n_j$ and $\mu_j$ is defined in \eqref{eqn:pbd.pa}. Thus, the performance of the Breslow and Efron estimators depends on the average $\bar{p}_{j}$ and the variation $\sum_{i\in \R(t_{(j)})}(p_{ij}-\bar{p}_{j})^2/n_j$ of the risk scores $r_i=\exp(\xvec_i^{\text{T}}\truebeta)$. The average and variation of risk scores depend on the values of $\truebeta$ and the distribution of the covariates $\xvec_i$. Thus both the scale of $\truebeta$ and the variation in  $\xvec_i$’s can affect the approximation accuracy.
\end{remark}

%%%%%%%%%%%%%%%%%%%%%%%%%%%%%%%%%%%%%%%%%%%%%%%%%%%%%%%%%%%%%%%%%%%%%%%%%%%%%%%%%
\subsection{Asymptotics under the Grouped Continuous Failure Time Model}\label{sec:theory_grp}
%%%%%%%%%%%%%%%%%%%%%%%%%%%%%%%%%%%%%%%%%%%%%%%%%%%%%%%%%%%%%%%%%%%%%%%%%%%%%%%%%

Now we study the large sample properties of the estimators of $\truebeta$ under the grouped continuous failure time model where event times can be tied. Among the estimators from the existing methods, we select the Breslow estimator $\betahatB$ to show its consistency and asymptotic normality, with the notion that the properties for the other estimators can be derived similarly, given our result in Theorem~\ref{thm:relation.to.pb}. Then we establish similar large sample properties for the PB distribution estimator $\betahatPB$ based on \eqref{eqn:PB_betahat}.

In this section, we restrict $t \in [0, \zeta]$. We denote by $\D(t) \equiv \{i: T_i^\ast=t\text{ and }\delta_i=1 \}$ and $\R(t)\equiv \{i: T_i^\ast \geq t\}$ respectively the event and at-risk sets at time $t$. Let $d(t)\equiv|\D(t)|$ and $n(t)\equiv|\R(t)|$. Let $\Omega_t=\Omega \cap [0,t]$ for some $t \in [0, \zeta]$. We define $\Delta H(t)=H(t)-H(t^-)$ for a function $H(\cdot)$. Let the PB distribution partial likelihood component at $t$ be 
$$L_t(\truebeta)=\frac{A_t(\truebeta)}{B_t(\truebeta)}=\frac{\left\{\prod_{i\in \D(t)} p_i(\truebeta,t)\right\}\left\{\prod_{i\in \R(t) \setminus \D(t)} \left(1- p_i(\truebeta,t)\right)\right\}}{\sum_{\A_{d(t)}\in\mathscr{F}_{d(t)}} \left\{\prod_{i\in \A_{d(t)}}{p_i(\truebeta,t)}\right\}\left\{\prod_{i\in \R(t) \setminus \A_{d(t)}}\left(1- p_i(\truebeta,t)\right) \right\}},$$ where $p_i(\truebeta,t)=1-\exp(-\exp( \xvec_i^{\text{T}} \truebeta ) \Delta\Lambdahat(t) )$. Let the Breslow partial likelihood component at $t$ be
$$L^{\text{b}}_t(\truebeta)=\frac{\exp\left(\sum_{i\in\D(t)}\xvec_{i}^{\text{T}}\truebeta\right)}{\left\{\sum_{i\in\R(t)}\exp(\xvec_{i}^{\text{T}}\truebeta)\right\}^{d(t)}/d(t)!}.$$ Here $\Lambdahat(t)= \sum_{j=1}^k \lambdahat_{j} \mathds{1}(t_{(j)} \leq t)=\int_0^td\Lambdahat(s)=\sum_{s \in \Omega_t}\Delta\Lambdahat(s)$ stands for any estimator of $\Lambda_0^{\ast}(t)$, which corresponds to $\lambdavechat$ in \eqref{eqn:PL.PB}. We write the PB distribution partial likelihood in \eqref{eqn:PL.PB} at $t \in [0, \zeta]$ as $L(\truebeta, \lambdavechat,t)=\prod_{s \in \Omega_t}L_s(\truebeta)$ and the Breslow partial likelihood in \eqref{eqn:breslow} at $t \in [0, \zeta]$ as $L^{\text{b}}(\truebeta,t) =\prod_{s \in \Omega_t}L_s^{\text{b}}(\truebeta)$, where if $t\notin \Omega^{\mathcal{K}}$ (i.e., $d(t)=0$), we have $\Delta \Lambdahat(t)=0$, $p_i(\truebeta,t)=0$, and $L^{\text{b}}_t(\truebeta)=L_t(\truebeta)=1$.

{\color{black}Let $\mathcal{B} \subseteq \mathbb{R}^d$ be an open neighborhood of $\truebeta$. Without loss of generality, we assume $\mathcal{B}$ is sufficiently large and $\betahatB,\betahatPB \in \mathcal{B}$, resulting in $\betahatB\equiv\argmax_{\tilde\truebeta\in\mathbb{R}^d} L^{\text{b}}(\tilde\truebeta,\zeta)=\argmax_{\tilde\truebeta\in\mathcal{B}} L^{\text{b}}(\tilde\truebeta,\zeta)$ and $\betahatPB\equiv\argmax_{\tilde\truebeta\in\mathbb{R}^d} L(\tilde\truebeta, \lambdavechat,$ $\zeta)=\argmax_{\tilde\truebeta\in\mathcal{B}} L(\tilde\truebeta, \lambdavechat,\zeta)$.}

For the counting processes, let $\tilde{N}_i(t)=\mathds{1}(\tilde{T}^{\ast}_i \leq t)$, $N_i(t)=\mathds{1}(T^{\ast}_i \leq t, \delta_i=1)=\mathds{1}(\tilde{T}^{\ast}_i \leq t, \tilde{T}^{\ast}_i \leq C^{\ast}_i)$, and $Y_i(t)=\mathds{1}(T^{\ast}_i \geq t)=\mathds{1}(\tilde{T}^{\ast}_i \geq t, C^{\ast}_i\geq t)$ be the underlying counting process, the observed counting process, and the at-risk process for the grouped continuous failure time model, respectively.
Let the history (filtration) for the grouped continuous failure time model be
\begin{align*}
  &\mathcal{F}_t=\sigma\left\{ N_i(u), Y_i(u^+), \xvec_i ; i=1,\dots,n ; 0 \leq u \leq t \right\} \: \text{and}\\
  &\mathcal{F}_{t^-}=\sigma\left\{ N_i(u), Y_i(u), \xvec_i  ; i=1,\dots,n ; 0 \leq u < t \right\}.
\end{align*}
Note that $dN_i(t)=Y_i(t) d\tilde{N}_i(t)$. Thus, one can show that $\E(dN_i(t)|\mathcal{F}_{t^-})=\pr(dN_i(t)=1|\mathcal{F}_{t^-})=Y_i(t)d\Lambda^{\ast}(t;\xvec_i)$ due to our assumption that $\{\tilde T_i^{\ast}\}_{i=1}^n$ and $\{C_i^{\ast}\}_{i=1}^n$ are conditionally independent given $\{\xvec_i\}_{i=1}^n$. We refer to equation (5.7) from the \citet{KalbfleischPrentice2002} for more details. So by the Doob-Meyer decomposition, $N_i(t)=A_i(t)+M_i(t)$, where $A_i(t)=\int_0^t Y_i(s) d\Lambda^{\ast}(s;\xvec_i)=\sum_{s \in \Omega_t}Y_i(s) \Delta \Lambda^{\ast}(s;\xvec_i)$ is the compensator and $M_i(t)=N_i(t)-A_i(t)$ is a zero-mean martingale. Let $N(t)=\sum_{i=1}^nN_i(t)$, $A(t)=\sum_{i=1}^nA_i(t)$, and $M(t)=\sum_{i=1}^nM_i(t)$. Again, we refer to Sections 5.2 and 5.3 of \citet{KalbfleischPrentice2002} for more details on the construction of counting processes and martingales.

Although the time points in $\Omega$ all depend on the grouping parameter $\tau$, for the simplicity of notation, we don't include $\tau$ into the subscripts. Every process or statement depending on a time point in $\Omega$ also depends on $\tau$ implicitly. Furthermore, one can see that $\tilde{T}_i^{\ast}$, $C_i^{\ast}$, $T_i^{\ast}$, and all their related functions {\color{black}including $\tilde{N}_i(t)$, $N_i(t)$, and $Y_i(t)$} depend on $\tau$. All our asymptotic notations and technical assumptions are collected in Appendices~\ref{Sec:asymp_notation} and~\ref{sec:assumptions}, respectively.

For $t \in [0, \zeta]$, the logarithm of the Breslow partial likelihood is
\begin{align}
    \log \left(L^{\text{b}}(\truebeta, t) \right)
    &=\sum_{i=1}^n \int_0^t\left\{ \xvec_i^{\text{T}}\truebeta-\log\left( \sum_{l=1}^nY_l(s) \exp( \xvec_l^{\text{T}} \truebeta) \right) \right\}dN_i(s)\\
    & \quad + \sum_{s \in \Omega_t} \log(\Delta N(s)!)\\
    &=\sum_{i=1}^n \sum_{s \in \Omega_t}\left\{ \xvec_i^{\text{T}}\truebeta-\log\left( \sum_{l=1}^nY_l(s) \exp( \xvec_l^{\text{T}} \truebeta) \right) \right\}\Delta N_i(s)\\& \quad + \sum_{s \in \Omega_t} \log(\Delta N(s)!).
\end{align}The score function is \begin{align}
U_{\text{b}}(\truebeta,t)=\frac{\partial}{\partial \truebeta} \log (L^{\text{b}}(\truebeta, t)) &=\sum_{i=1}^n \int_0^t\left( \xvec_i - \epsilon(\truebeta, s) \right)dN_i(s)\\&=\sum_{i=1}^n \sum_{s \in \Omega_t}\left( \xvec_i - \epsilon(\truebeta, s) \right)\Delta N_i(s),    
\end{align}where $\epsilon(\truebeta, t)=\sum_{i=1}^n q_i(\truebeta, t) \xvec_i$ with $q_i(\truebeta, t)=Y_i(t) \exp (\xvec_i^{\text{T}} \truebeta) / ( \sum_{l=1}^n Y_l(t)$ $ \exp(\xvec_l^{\text{T}} \truebeta) )$, which satisfies $\sum_{i=1}^n q_i(\truebeta, t)=1$. The information matrix is $$I_{\text{b}}(\truebeta,t)=-\frac{\partial^2}{\partial \truebeta \partial \truebeta^{\text{T}}}\log ( L^{\text{b}}(\truebeta, t) )=\int_0^t \mathcal{V}(\truebeta, s) dN(s)
    =\sum_{s \in \Omega_t} \mathcal{V}(\truebeta, s) \Delta N(s),$$
 where $\mathcal{V}(\truebeta, t)=\sum_{i=1}^n q_i(\truebeta, t) ( \xvec_i-\epsilon(\truebeta, t)) ( \xvec_i-\epsilon(\truebeta, t))^{\text{T}} $. Let $\Sigma(\truebeta, \zeta)$ be the matrix defined in Assumption~\ref{assump:grp_consistency}. Our asymptotic results for $\betahatB$ are
\begin{thm}\label{thm:consistent.grouped}
Under Assumptions \ref{assump:enough_sample_cmg} -- \ref{assump:grp_consistency}, $\betahatB$ converges in probability to $\truebeta$ and $I_{\text{b}}(\betahatB, \zeta)/n$ converges in probability to $\Sigma(\truebeta, \zeta)$
when $\tau \rightarrow 0$ as $n \rightarrow \infty$.
\end{thm}\begin{thm}\label{thm:clt.grouped}
Under Assumptions \ref{assump:enough_sample_cmg} -- \ref{assump:grp_consistency}, $n^{-1/2} U_{\text{b}}(\truebeta, \zeta)$ converges in distribution to $ \text{N}\left(\boldsymbol{0}, \Sigma(\truebeta, \zeta)\right)$ and $n^{1/2}( \betahatB - \truebeta) $ converges in distribution to $ \text{N}\left(\boldsymbol{0}, \Sigma(\truebeta, \zeta)^{-1}\right)$ when $n^{1/2}\tau \rightarrow 0$ as $n \rightarrow \infty$.\end{thm}

The proofs of Theorems \ref{thm:consistent.grouped} and \ref{thm:clt.grouped} are respectively in Supplementary Sections S1.6 and S1.7. As we have $\sup_{t \in \Omega}d(t)=\bigO_{\text{P}}(n\tau)$ for $t \in \Omega$ under Assumption \ref{assump:tie_order}, if the order of $\tau$ is no smaller than $1/n$ and $\tau \rightarrow 0$, the Breslow method can achieve consistency allowing ties. If the order of $\tau$ is no smaller than $1/n$ and $n^{1/2}\tau \rightarrow 0$, the Breslow estimator have the asymptotic normality allowing ties.

Now we derive the asymptotic properties of the PB distribution estimator with three more assumptions, \ref{assump:order_tau_n} -- \ref{assump:grp_pbd_an}. In particular, Assumption~\ref{assump:order_tau_n} requires $\tau$ to be of the order $1/n$. The following theorem shows the asymptotic equivalence between $L(\tilde\truebeta, \lambdavechat,t)$ and $L^{\text{b}}(\tilde\truebeta,t)$ {\color{black}uniformly} for $\tilde\truebeta \in \mathcal{B}$.{\color{black}\begin{thm}\label{thm:PL.converge}
When Assumptions \ref{assump:enough_sample_cmg} -- \ref{assump:tie_order} and \ref{assump:order_tau_n} -- \ref{assump:efron} are satisfied, for all $t \in [0,\zeta]$, $$\sup_{\tilde{\truebeta}\in \mathcal{B}}\left|\log \left( L(\tilde\truebeta, \lambdavechat,t) \right)-\log \left( L^{\text{b}}(\tilde\truebeta,t) \right) \right|= \bigO_{\text{P}}(1).$$
\end{thm}}

Let $X_{\text{pb}}(\tilde\truebeta,\lambdavechat,t)=\{\log (L(\tilde\truebeta,\lambdavechat,t))-\log (L(\truebeta,\lambdavechat,t))\}/n$ and $ X(\tilde\truebeta, t) =\{\log (L^{\text{b}}(\tilde\truebeta,t))-\log( L^{\text{b}}(\truebeta, t))\}/n$. By Theorem \ref{thm:PL.converge}, when Assumptions \ref{assump:enough_sample_cmg} -- \ref{assump:tie_order}  and \ref{assump:order_tau_n} -- \ref{assump:efron} are satisfied, {\color{black}we have $\sup_{\tilde\truebeta \in \mathcal{B}}|X_{\text{pb}}(\tilde\truebeta,\lambdavechat,t)-X(\tilde\truebeta,t)|=\bigO_{\text{P}}(1/n)$ for all $t \in [0, \zeta]$,} where $\betahatPB= \argmax_{\tilde\truebeta\in \mathcal B}X_{\text{pb}}(\tilde\truebeta,\lambdavechat, \zeta)$ and $\betahatB= \argmax_{\tilde\truebeta\in \mathcal B}X(\tilde\truebeta,$ $ \zeta)$. Based on closeness between PB distribution and Breslow partial likelihoods by Theorem \ref{thm:PL.converge}, the asymptotic results for $\betahatPB$ are

\begin{thm}\label{thm:pb.consistent.grouped}
Under Assumptions \ref{assump:enough_sample_cmg} -- \ref{assump:efron}, $\betahatPB $ converges in probability to $ \truebeta$ and $I_{\text{b}}(\betahatPB, \zeta)/n $ converges in probability to $ \Sigma(\truebeta, \zeta)$ as $n \rightarrow \infty$.
\end{thm}
\begin{thm}\label{thm:pb.clt.grouped}
Under Assumptions \ref{assump:enough_sample_cmg} -- \ref{assump:grp_pbd_an}, the quantity $n^{1/2}( \betahatPB - \truebeta) $ converges in distribution to $\text{N}\left(\boldsymbol{0}, \Sigma(\truebeta, \zeta)^{-1}\right)$ as $n \rightarrow \infty$.
\end{thm}

The proofs of Theorems  \ref{thm:pb.consistent.grouped} and \ref{thm:pb.clt.grouped} are respectively in Supplementary Sections S1.10 and S1.11. Even though the estimators are all asymptotically unbiased, the score function based on the Breslow and Efron methods do not have a zero mean due to the Poisson approximation under small samples. However, the accurate likelihood calculation based on the PB distribution provides the exact probability under any sample size. Thus, the score function always has a zero mean using the PB distribution calculation if the true $\lambdavec$ is used. That is, although the new method uses $\lambdavechat$ instead of $\lambdavec$, it tends to have less bias even under small samples. We indeed observe a smaller bias for the new estimator in small samples from the simulation studies in Section \ref{sec:simul}. Because of the variation caused by $\lambdavechat$, the PB distribution estimator tends to have a slightly larger variance in practice. We also observe this in the simulation studies. However, the difference is minimal.

%%%%%%%%%%%%%%%%%%%%%%%%%%%%%%%%%%%%%%%%%%%%%%%%%%%%%%%%%%%%%%%%%%%%%%%%%%%%%%%%%
\subsection{Asymptotics Under Continuous Model}\label{sec:theory_conti}
%%%%%%%%%%%%%%%%%%%%%%%%%%%%%%%%%%%%%%%%%%%%%%%%%%%%%%%%%%%%%%%%%%%%%%%%%%%%%%%%%
In this section we show that the PB distribution estimator $\betahatPB$ based on $\eqref{eqn:PB_betahat}$ still possesses excellent asymptotic properties even when observations are generated from a continuous model and contain no ties. 
Let $L_j^{\text{c}}(\tilde\truebeta)={\exp(\xvec_{j_1}^{\text{T}}\tilde\truebeta)}$ $/\{\sum_{i\in\R(t_{(j)})}\exp(\xvec_{i}^{\text{T}}\tilde\truebeta)\},$ $\loglik_{\text{c}}(\tilde\truebeta)=\sum_{j=1}^k \log(L_j^{\text{c}}(\tilde\truebeta))$, $U_{\text{c}}(\tilde\truebeta, \zeta)=\partial\loglik_{\text{c}}(\tilde\truebeta)/(\partial\tilde\truebeta)$, and $I_{\text{c}}(\tilde\truebeta, \zeta)=-\partial^2\loglik_{\text{c}}(\tilde\truebeta)/(\partial\tilde\truebeta\partial\tilde\truebeta^{\text{T}})$ be respectively the log partial likelihood, the score function, and the information matrix for the approximate partial likelihood in \eqref{eqn:PL.CA.j}. Let $\widehat{\truebeta}_{\text{c}}\equiv\argmax_{\tilde{\truebeta}\in \mathbb{R}^d}\loglik_{\text{c}}(\tilde\truebeta)$ the estimator based on \eqref{eqn:PL.CA.j}. Without loss of generality, assume $\widehat{\truebeta}_{\text{c}}\in\mathcal{B}$. So we can write $\widehat{\truebeta}_{\text{c}}$ as $\argmax_{\tilde{\truebeta}\in \mathcal{B}}\loglik_{\text{c}}(\tilde\truebeta)$. Let $\loglik(\tilde\truebeta, \lambdavechat)=\log(L(\tilde\truebeta, \lambdavechat))$, $\chi_{\text{pb}}(\tilde\truebeta,\lambdavechat)=\{\loglik(\tilde\truebeta,\lambdavechat)-\loglik(\truebeta,\lambdavechat)\}/n$, and $\chi(\tilde\truebeta)=\{\loglik_{\text{c}}(\tilde\truebeta)-\loglik_{\text{c}}(\truebeta)\}/n$. One can see that the estimators {\color{black}satisfy} $\betahatPB= \argmax_{\tilde\truebeta\in \mathcal B}\chi_{\text{pb}}(\tilde\truebeta,\lambdavechat)$ and $\widehat{\truebeta}_{\text{c}}= \argmax_{\tilde\truebeta\in \mathcal B}\chi(\tilde\truebeta)$. Let $\Sigma_{\text{c}}(\truebeta, \zeta)$ be the matrix defined in Assumption~\ref{assump:conti_consistency}. The asymptotic results for the PB distribution estimator $\betahatPB$ under the continuous model without ties are

\begin{thm}\label{thm:consistent.continuous}
Under Assumptions \ref{assump:conti_no_out_lier} and \ref{assump:conti_consistency}, $\betahatPB $ converges in probability to $ \truebeta$ and $I_{\text{c}}(\betahatPB, \zeta)/n $ converges in probability to $ \Sigma_{\text{c}}(\truebeta, \zeta)$ as $n \rightarrow \infty$.
\end{thm}
\begin{thm}\label{thm:clt.continuous}
Under Assumptions \ref{assump:conti_no_out_lier} and  \ref{assump:conti_consistency} -- \ref{assump:conti_pbd_an}, $n^{1/2}( \betahatPB - \truebeta ) $ converges in distribution to $ \text{N}\left(\boldsymbol{0}, \Sigma_{\text{c}}(\truebeta, \zeta)^{-1}\right)$ as $n \rightarrow \infty$.
\end{thm}

The proofs of Theorems \ref{thm:consistent.continuous} and \ref{thm:clt.continuous}, similar to those of Theorems \ref{thm:pb.consistent.grouped} and \ref{thm:pb.clt.grouped}, are in Supplementary Section S1.12.

%%%%%%%%%%%%%%%%%%%%%%%%%%%%%%%%%%%%%%%%%%%%%%%%%%%%%%%%%%%%%%%%%%%%%%%%%%%%%%%%%
\section{Simulation Studies}\label{sec:simul}
%%%%%%%%%%%%%%%%%%%%%%%%%%%%%%%%%%%%%%%%%%%%%%%%%%%%%%%%%%%%%%%%%%%%%%%%%%%%%%%%%

Our simulation studies are done in settings with a single covariate \( x_i \), using scalar notation for simplicity, but it can be readily extended to multiple covariates. Overall, we follow the settings for the underlying continuous model described in Section \ref{sec:underlying_model}. For the covariate, we generate $n$ i.i.d. $x_i$’s from $\text{N}(0,\sigma_x^2)$. For event times, we use the Weibull distribution with the baseline hazard function $\lambda_0(t)=\gamma t^{\gamma-1}/\eta^\gamma$ such that
$\lambda(t;x_i)=\gamma \eta^{-\gamma}t^{\gamma-1}\exp(x_i\beta)=\gamma\eta_i^{-\gamma}t^{\gamma-1}$,
where $\eta_i=\eta \exp(-x_i \beta/\gamma)$. In particular, we generate $\tilde T_i$ as
$\tilde T_i=\exp(\mu_i+\sigma W_i)$,
where $\sigma=1/\gamma$, $\mu_i=\log(\eta_i)$, and $W_i$’s are $n$ i.i.d. following the standard smallest extreme value distribution.
For censoring times, we first generate $n$ i.i.d. $\tilde{C}_i$’s from the Weibull distribution with hazard function $\lambda_c(t)={\gamma_c}t^{\gamma_c-1}/{\eta_c^{\gamma_c}}$ and then set the censoring times as $C_i=\max\{\tilde{C}_i,\zeta\}$. To generate ties by grouping, we use the grouped continuous failure time model rule described in Section~\ref{sec:CMG} with several $\tau$ values.
We fix $\zeta=1$, $\eta=\eta_c=1.31$ and $\gamma=\gamma_c=1.5$, and repeat the simulations $B=10,000$ times for all the simulation cases. We vary $\beta \in \{1,1.5\}$, $\sigma_x \in \{1.5,2\}$, $\tau \in \{0.01, 0.1, 0.2\}$, and $n \in \{50,100,200,500,1000\}$.

For estimation performance metrics, we use the root mean square error (RMSE) and the absolute bias ($|\text{Bias}|$), each scaled by the true $\beta$, to compare our PB distribution method with the Breslow method and the Efron method. 
To evaluate inference performance, we compare the empirical coverage rates of confidence intervals and the average standard errors for the three methods with $n \in \{50,100,200\}$. The standard errors are calculated using respectively \( I_{\text{b}}^{-1/2}(\sbetahatB, \zeta) \) for the Breslow method, \( I_{\text{e}}^{-1/2}(\sbetahatE, \zeta) \) for the Efron method, and \( I_{\text{b}}^{-1/2}(\sbetahatPB, \zeta) \) for the PB distribution method.

\begin{figure}
    % 9 figures
    \centering
    \includegraphics[scale=0.25]{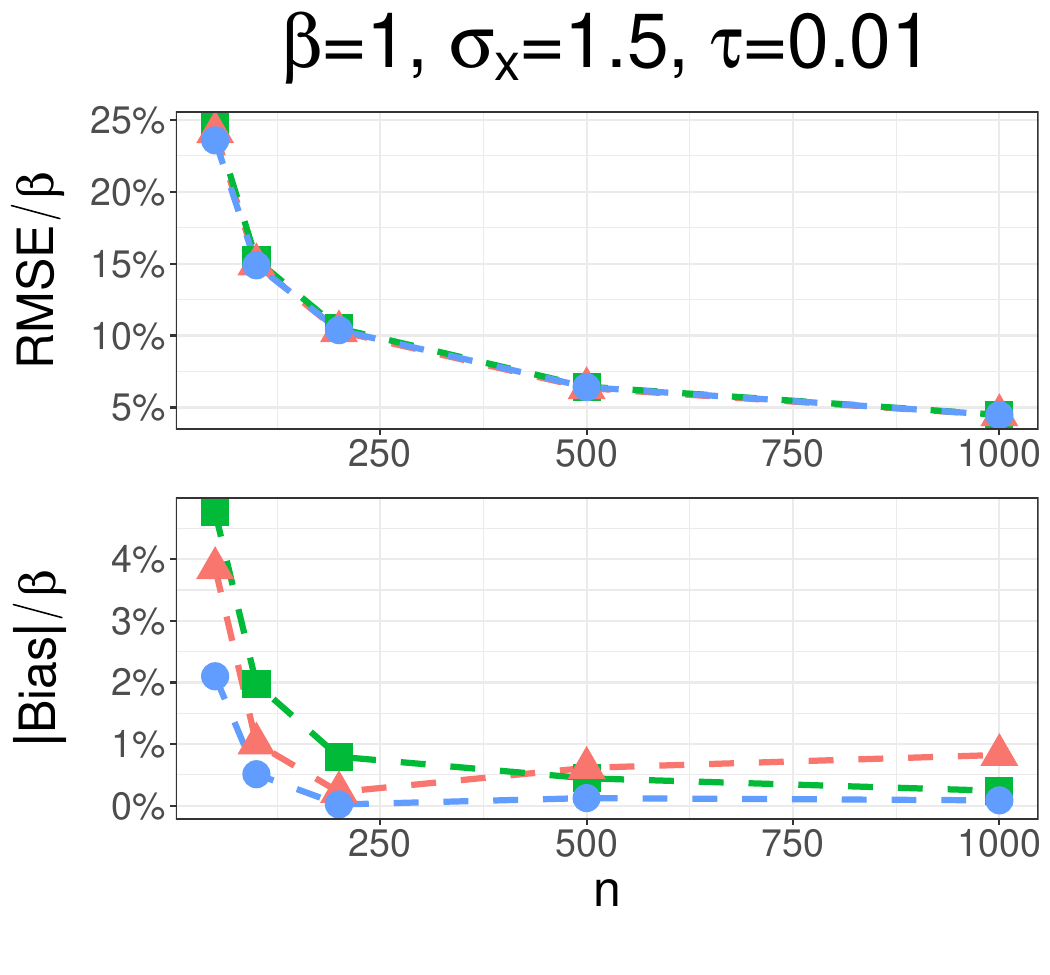}
    \includegraphics[scale=0.25]{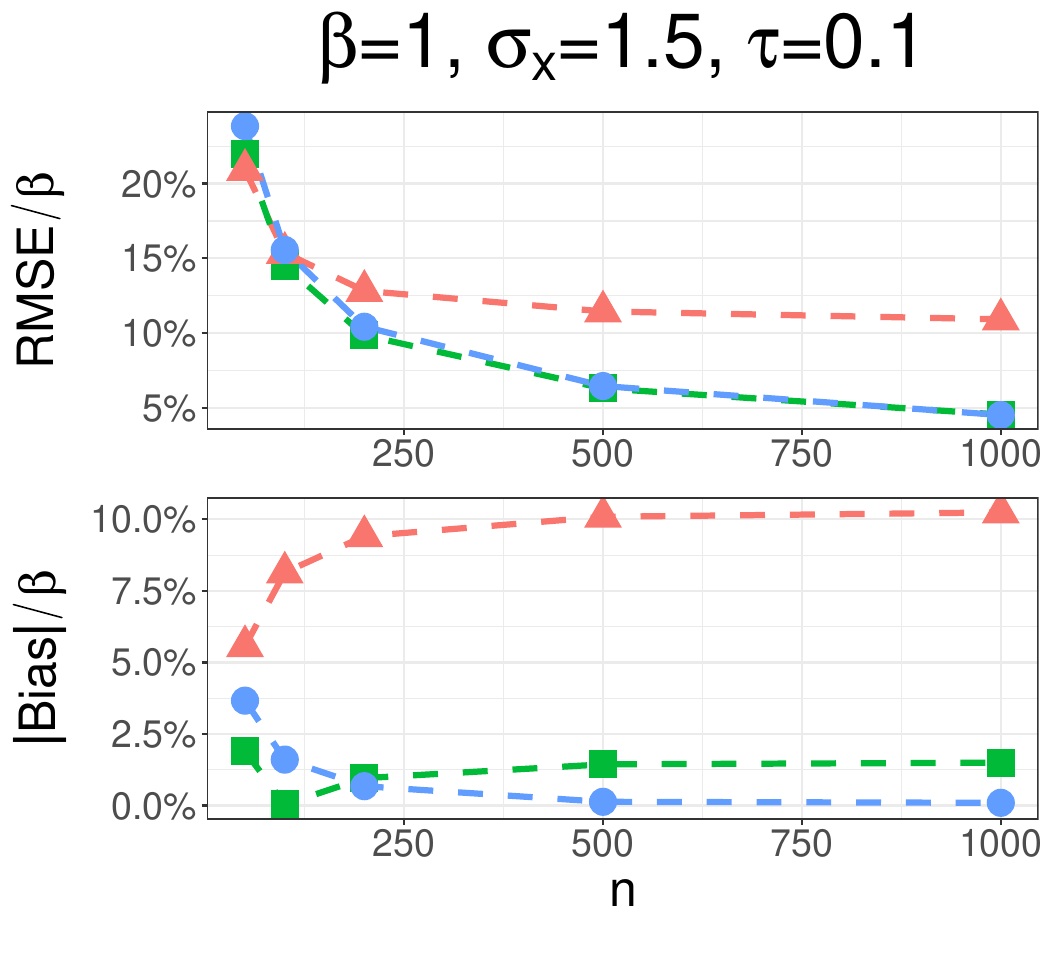}
    \includegraphics[scale=0.25]{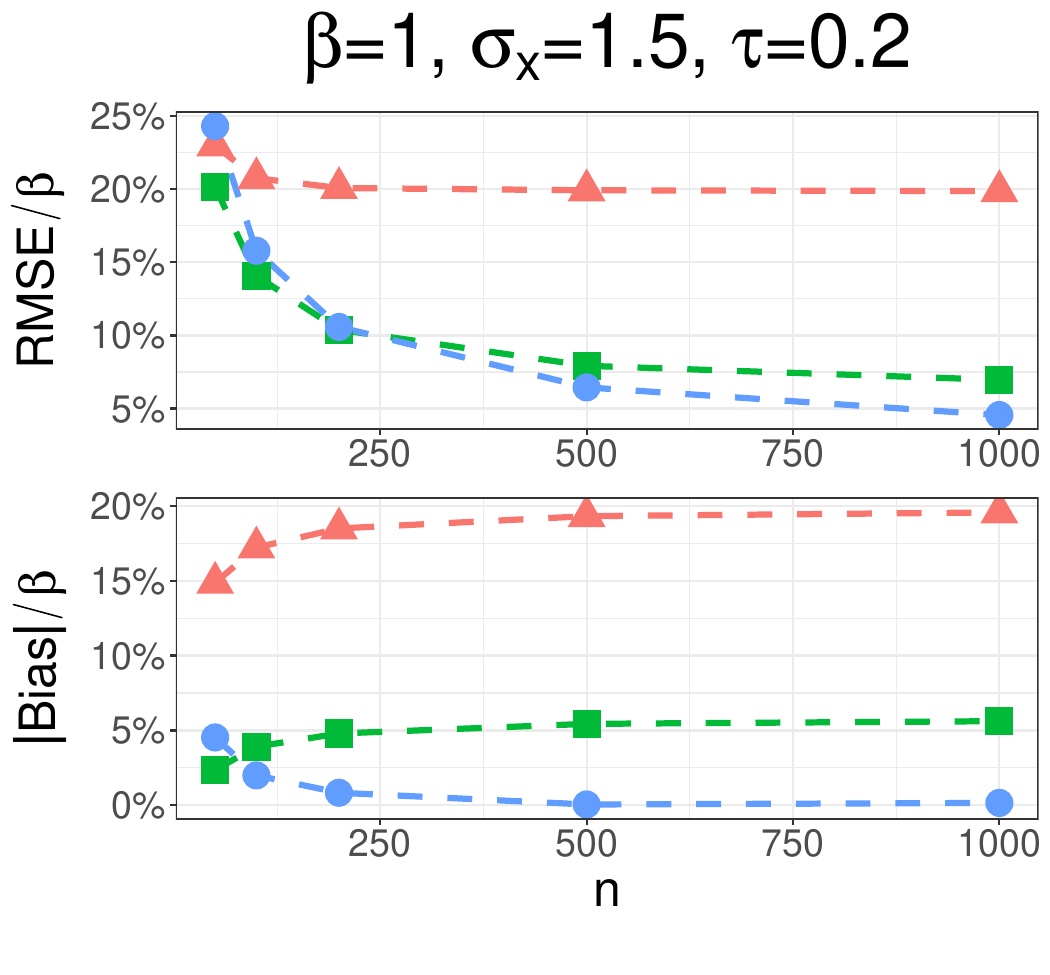}\\
    \includegraphics[scale=0.25]{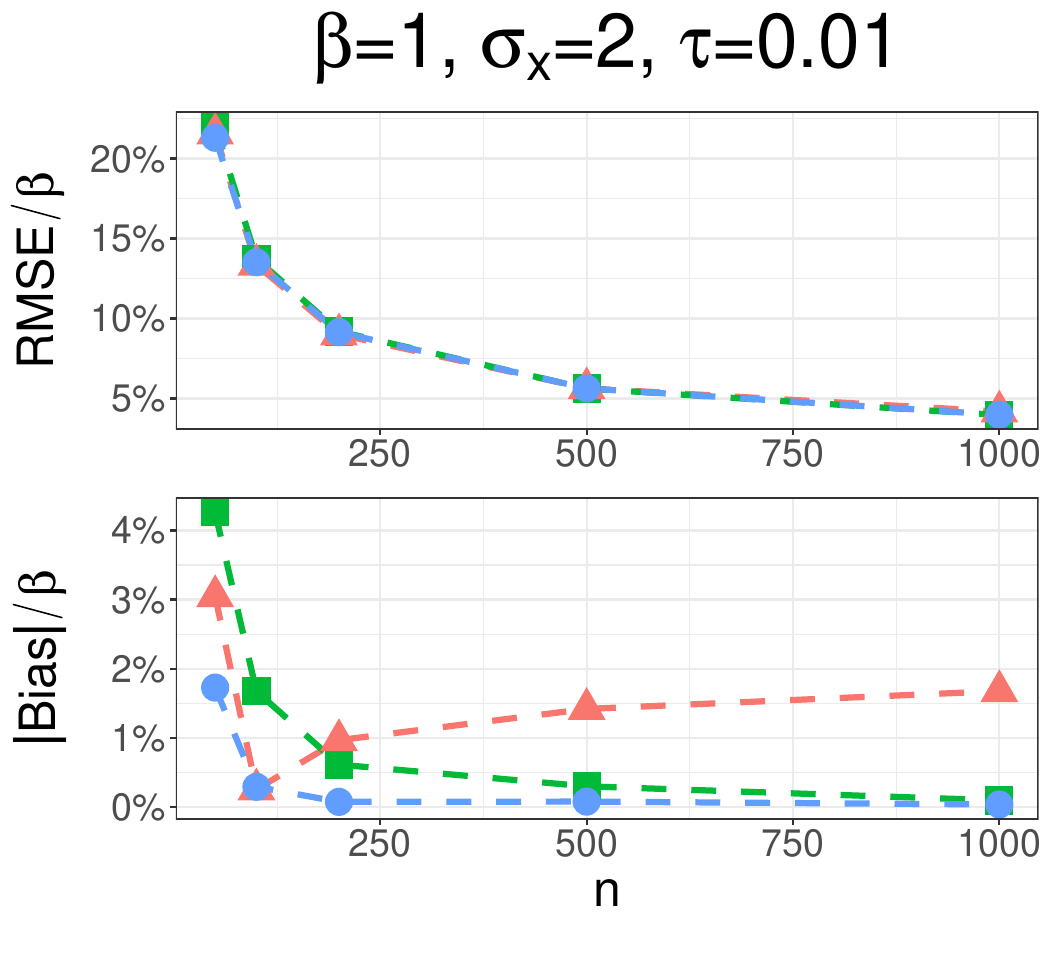}
    \includegraphics[scale=0.25]{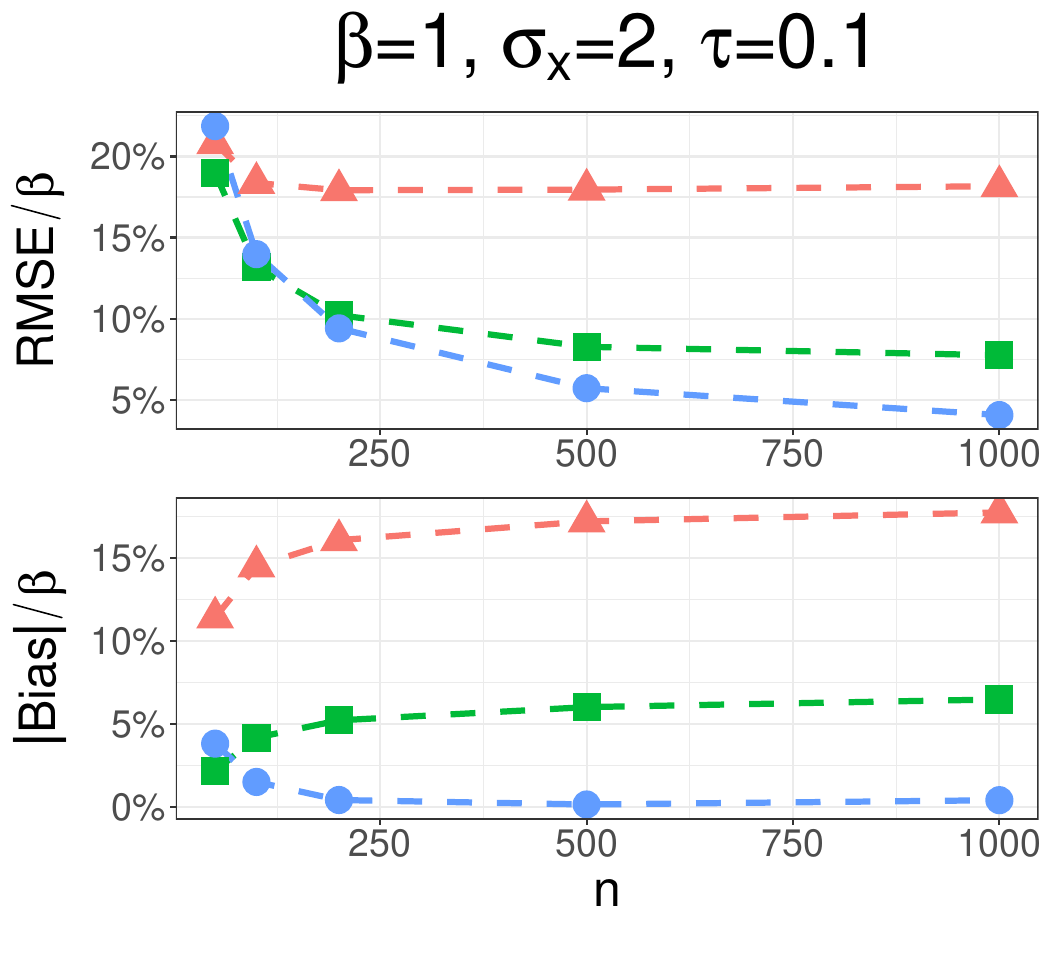}
    \includegraphics[scale=0.25]{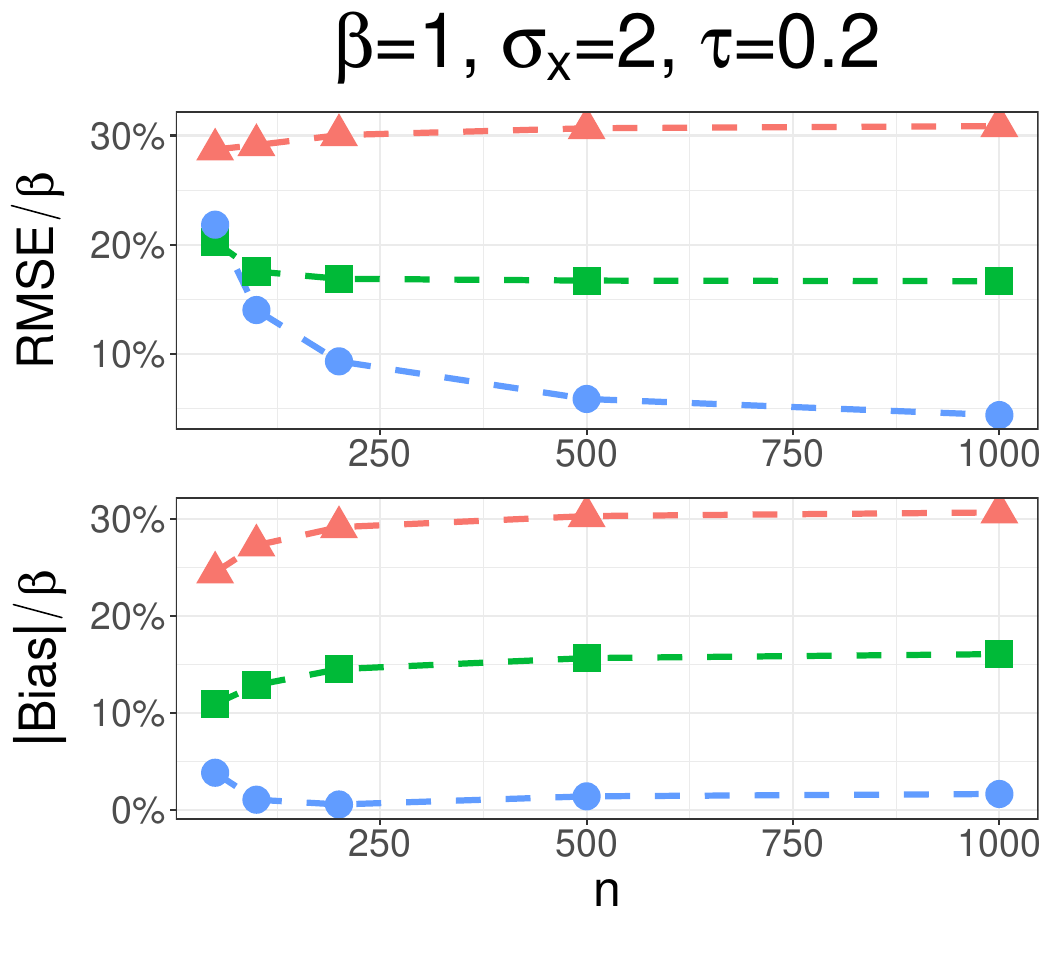}\\
    \includegraphics[scale=0.25]{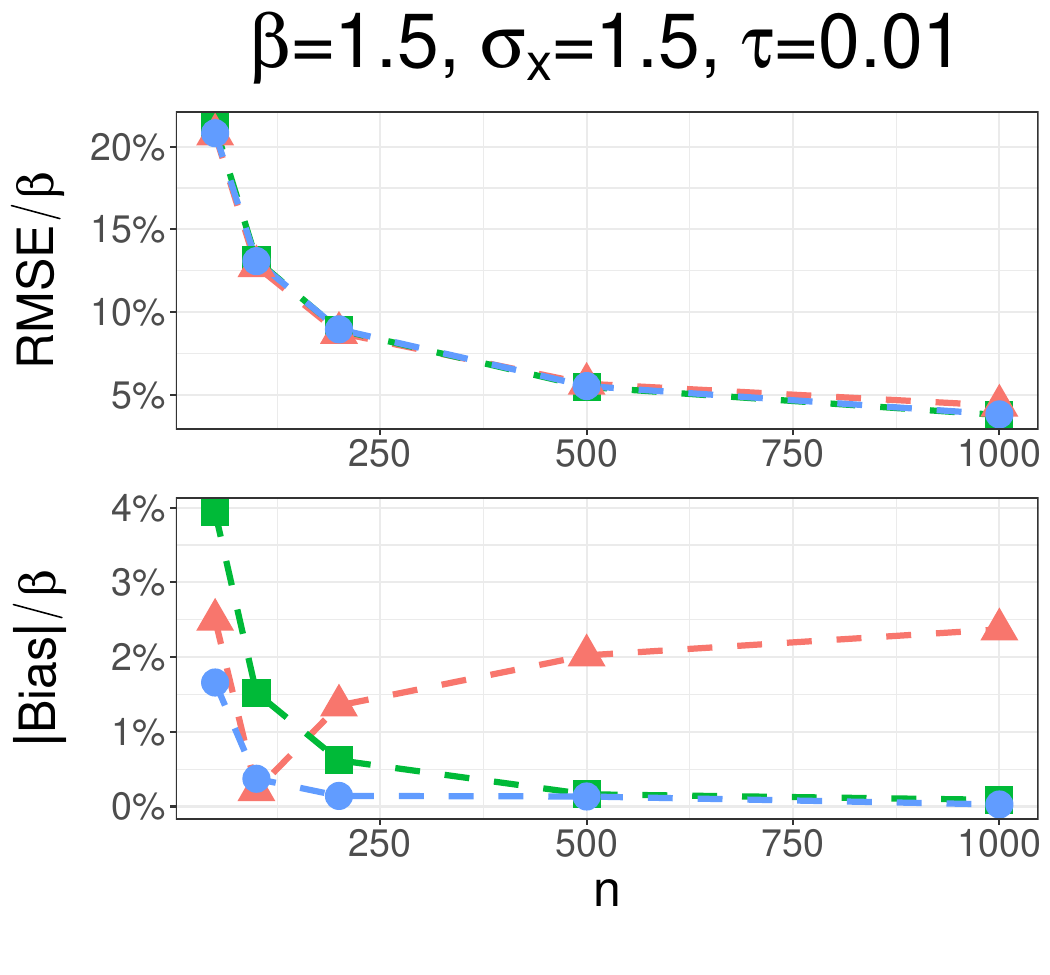}
    \includegraphics[scale=0.25]{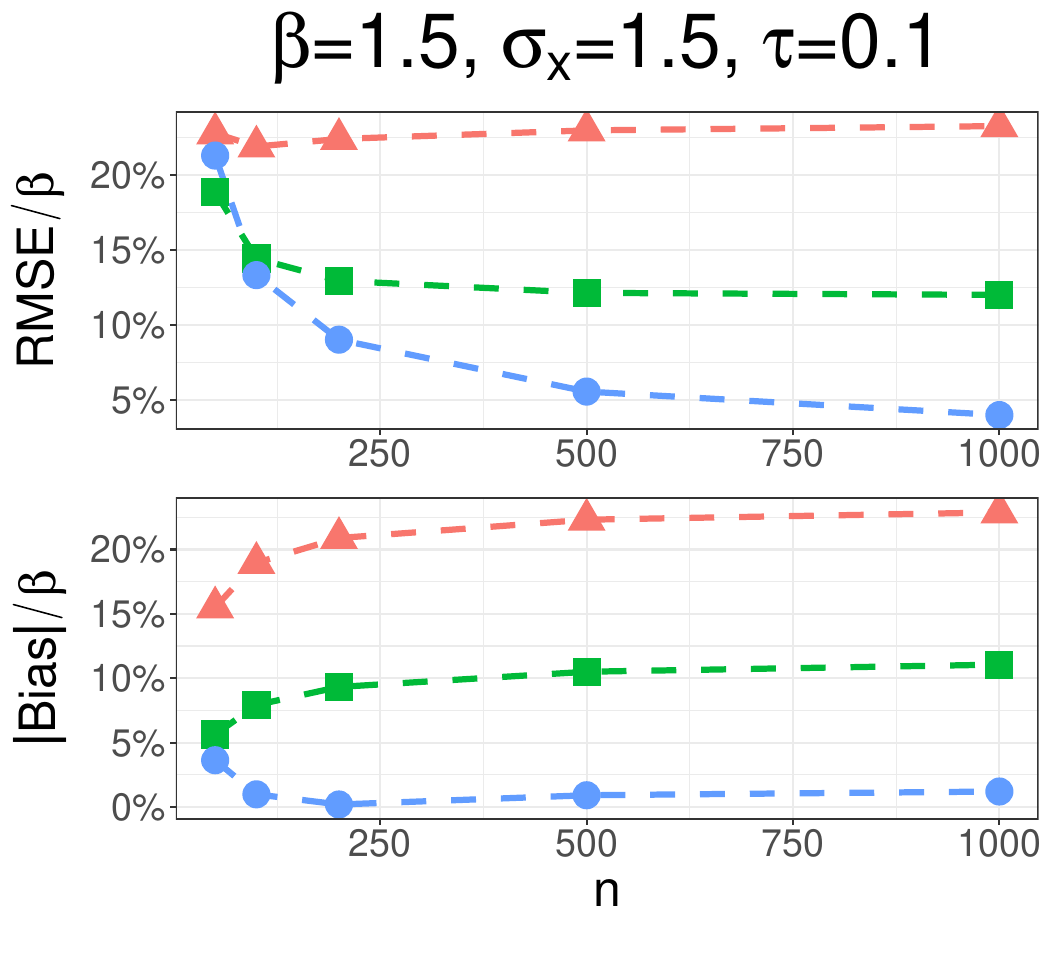}
    \includegraphics[scale=0.25]{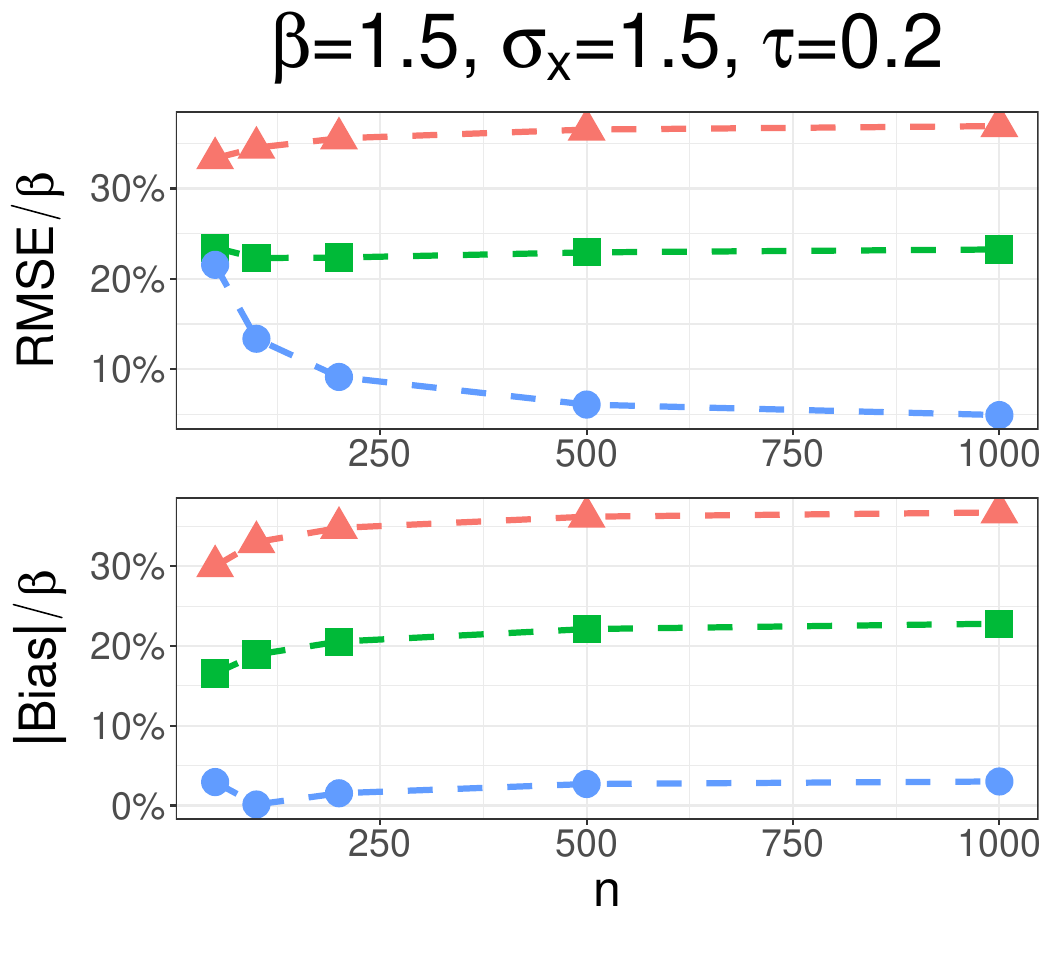}\\
    \includegraphics[scale=0.25]{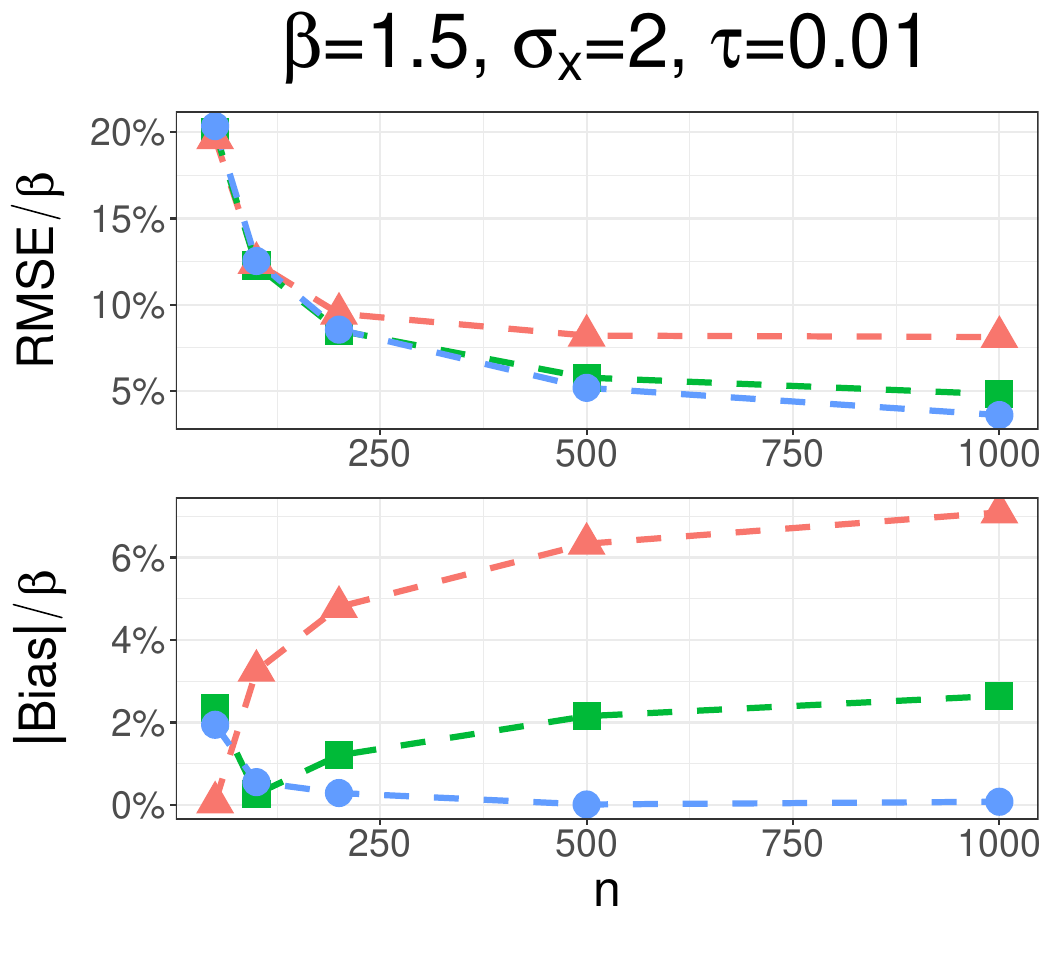}
    \includegraphics[scale=0.25]{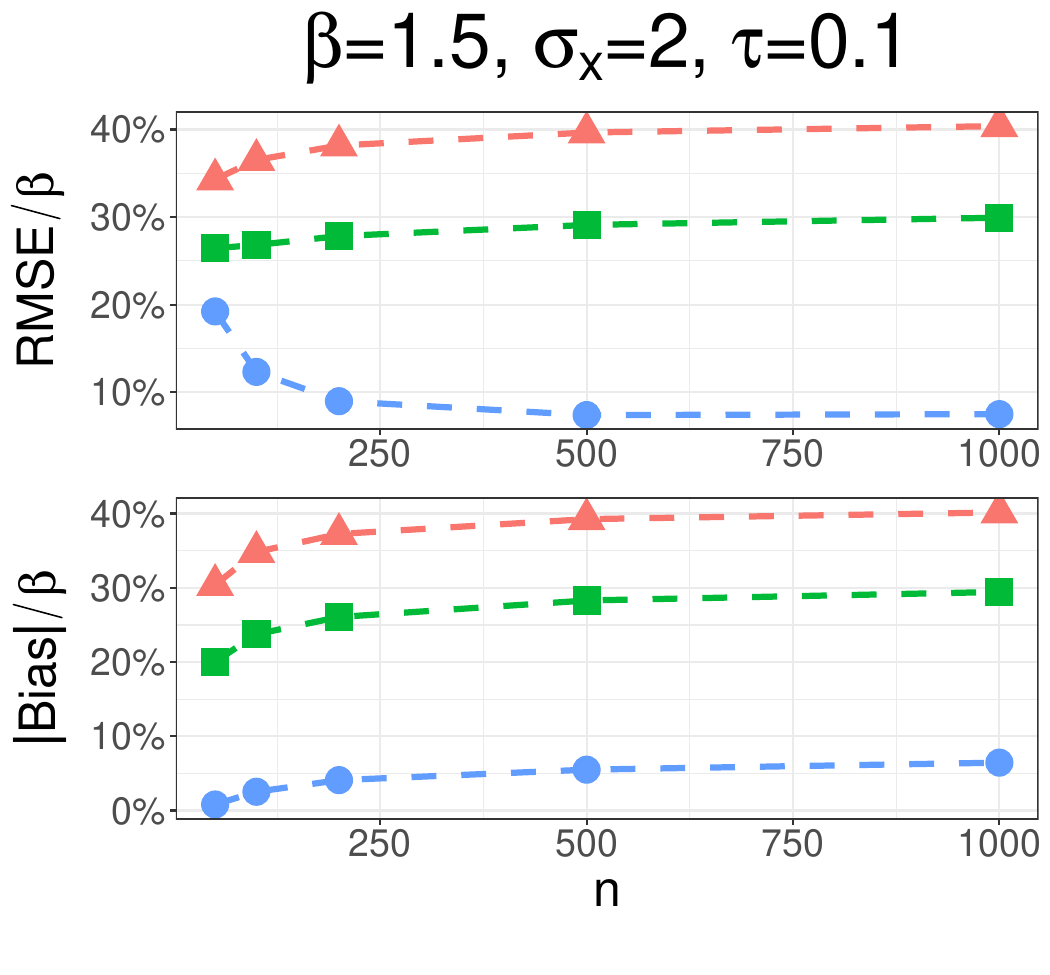}
    \includegraphics[scale=0.25]{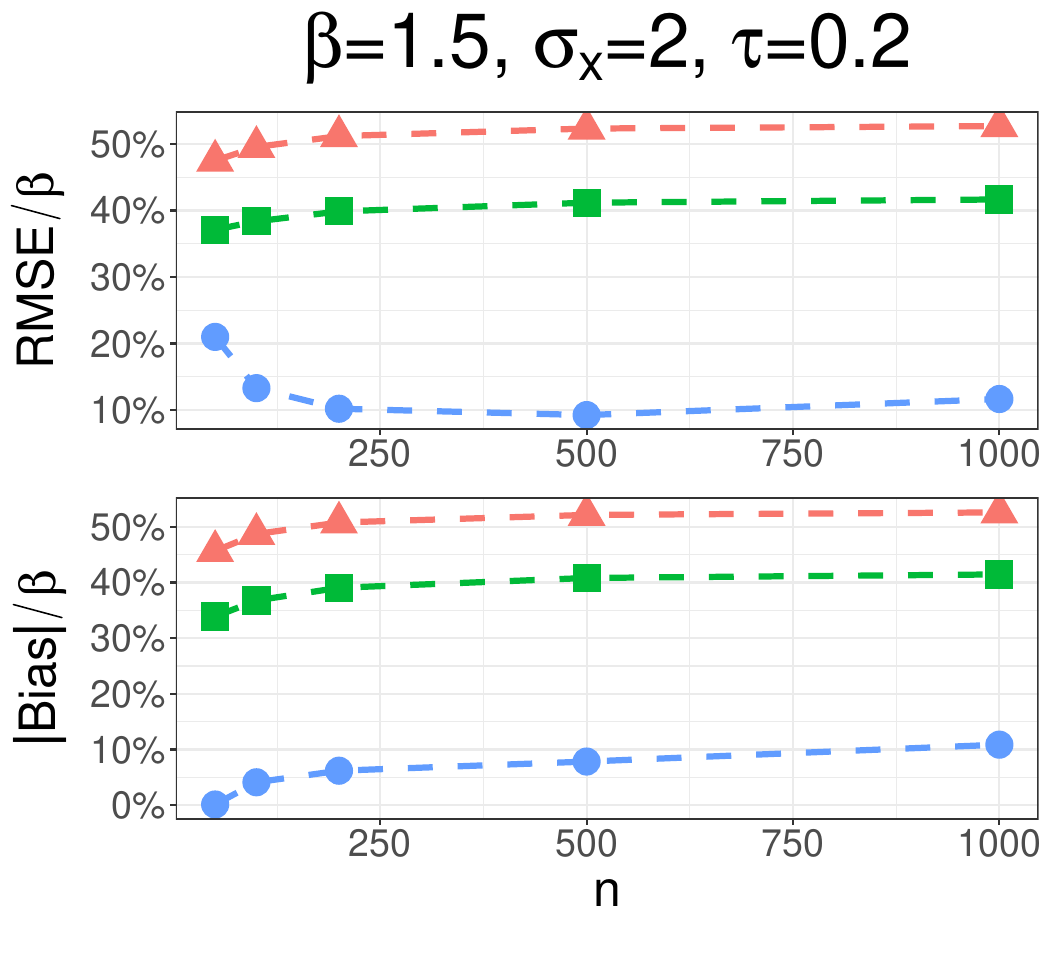}\\
    \includegraphics[scale=0.25]{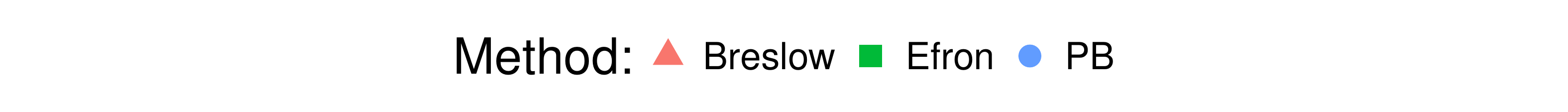}
    \caption{Simulation results for RMSE and $|\text{Bias}|$.}\label{Fig_beta_1_nor1}
\end{figure}

Figure~\ref{Fig_beta_1_nor1} illustrate that for many cases, $\sbetahatPB$ exhibits notably lower $|\text{Bias}|$ and smaller RMSE compared to existing methods. On the other hand, the standard deviations of the estimators from the three methods are similar, as shown in Supplementary Figure S1. This indicates that $\sbetahatPB$'s better RMSE performance primarily arises from its significant reduction in $|\text{Bias}|$. To be more specific, when $\beta, \sigma_{x}^2,$ and $\tau$ are all small, all three methods give comparable performance. When at least one of these model parameters gets bigger, the Breslow method's performance starts to deteriorate. The Efron method can maintain a performance competitive to the PB distribution method until at least two of these model parameters get bigger. Besides its dominance in the cases of larger $\tau$, $\sigma_x^2$, or $\beta$, the PB distribution estimators can deliver a competitive performance in all the other cases too.

Table~\ref{Table_CI} provides empirical coverage rates of confidence intervals and average standard errors. Clearly, the three methods deliver comparable standard errors. In terms of the empirical coverage, the message is similar to that of the esimation performance. When all the three model parameters, \(\tau\), \(\beta\), and covariate variation, are small, the three method all perform well with coverage rates close to the nominal. When at least one of the three model parameters gets larger, the coverage rates for the PB distribution method dominate those for the other methods.

Overall, our method outperforms the Breslow and Efron methods in both parameter estimation and confidence interval coverages when at least one of the three model parameters, \(\tau\), \(\beta\), or covariate variation, gets larger. Intuitively, an increase in $\sigma_x^2$ or $\beta$ leads to an increase in the variation among $p_{ij}$'s. As discussed in Remark~\ref{rmk:Lecam}, an increased level of variation among $p_{ij}$'s leads to a larger value of ${\sum_{i\in \R(t_{(j)})}p_{ij}^2}/n_j$, which can also be enlarged by increment in $\tau$ considering the fact that $\sup_{ij}p_{ij}=\bigO_{\text{P}}(\tau)$. This makes the Poisson approximation in the Breslow partial likelihood \eqref{eqn:breslow} and the Efron partial likelihood \eqref{eqn:efron} less precise, and thus reduces the accuracy of $\sbetahatB$ and $\sbetahatE$. Additionally, note that Theorem \ref{thm:PL.converge} requires Assumption \ref{assump:order_tau_n}, which is $\tau\asymp 1/n$. Or more simply, note that $\sup_{ij} |p_{ij}-\exp(x_i\beta)\lambda_j| =\bigO_{\text{P}}(\tau^2)$. Thus, an increment in $\tau$ makes \eqref{eqn:breslow} and \eqref{eqn:efron} less precise, and thus less accurate $\sbetahatB$ and $\sbetahatE$.

As a side note, the average computing time for our method with \(\beta=1.5\), \(\sigma_x=1.5\), \(\tau=0.2\), and \(n=200\) is approximately 0.028 seconds, on a MacBook with an 8-core Apple M1 chip and 17.2 GB RAM, echoing its computational efficiency demonstrated in \cite{Hong2013}.

\begin{table}
\caption{Simulation results for inference. Each cell shows the empirical coverage rate of the confidence interval and the average estimated standard error (in parentheses).\label{Table_CI}}
\begin{center}
%\scriptsize
\renewcommand{\arraystretch}{0.9} 
\begin{tabular}{ccccccc}
\hline
\multirow{2}{*}{$\beta$} & \multirow{2}{*}{$\sigma_x$} & \multirow{2}{*}{$\tau$}   & \multicolumn{1}{c}{\multirow{2}{*}{Method}}   & \multicolumn{3}{c}{$n$}                 \\\cline{5-7}
                 & \multicolumn{1}{c}{}                        &&      & 50    & 100   & 200    \\ \hline\hline
\multirow{18}{*}{1}&\multirow{9}{*}{1.5}&\multirow{3}{*}{0.01}& Breslow                                     & 0.959(0.226)& 0.956(0.150)& 0.952(0.102)\\ && & Efron                                        & 0.958(0.226)& 0.954(0.150)& 0.952(0.103)\\
                       
                     &&  & PB                                 & 0.957(0.224)& 0.956(0.149)& 0.952(0.103)\\
%                       & PBD-2                                 & 0.937(0.208)& 0.942(0.142)& 0.942(0.099)\\
\cline{3-7}
&&\multirow{3}{*}{0.1}& Breslow                                     & 0.939(0.211)& 0.903(0.140)& 0.828(0.096)\\  && & Efron                                       & 0.959(0.217)& 0.951(0.145)& 0.950(0.100)\\
                       
                     &&  & PB                                  & 0.953(0.219)& 0.939(0.145)& 0.942(0.099)\\
%                       & PBD-2                                & 0.926(0.202)& 0.920(0.136)& 0.928(0.094)\\
\cline{3-7}
&&\multirow{3}{*}{0.2}& Breslow                                     & 0.857(0.196)& 0.713(0.131)& 0.441(0.090)\\  && & Efron                                      & 0.949(0.205)& 0.934(0.138)& 0.908(0.095)\\
                       
                    &&   & PB                                  & 0.941(0.213)& 0.932(0.140)& 0.922(0.094)\\
%                       & PBD-2                                & 0.917(0.201)& 0.917(0.135)& 0.916(0.092)\\
\cline{2-7}

&\multirow{9}{*}{2}&\multirow{3}{*}{0.01}& Breslow                                     & 0.958(0.202)& 0.952(0.132)& 0.948(0.090)\\ && & Efron                                        & 0.958(0.203)& 0.953(0.133)& 0.954(0.091)\\
                       
                    &&   & PB                                 & 0.954(0.200)& 0.951(0.132)& 0.950(0.090)\\
%                       & PBD-2                                 & 0.926(0.180)& 0.930(0.122)& 0.932(0.085)\\
\cline{3-7}
&&\multirow{3}{*}{0.1}& Breslow                                     & 0.856(0.179)& 0.715(0.117)& 0.460(0.079)\\  && & Efron                                       & 0.936(0.185)& 0.910(0.123)& 0.864(0.084)\\
                       
                    &&   & PB                                  & 0.950(0.197)& 0.937(0.128)& 0.931(0.086)\\
%                       & PBD-2                                & 0.904(0.173)& 0.905(0.115)& 0.905(0.079)\\
\cline{3-7}
&&\multirow{3}{*}{0.2}& Breslow                                     & 0.592(0.158)& 0.282(0.104)& 0.041(0.071)\\  && & Efron                                      & 0.835(0.166)& 0.712(0.110)& 0.498(0.075)\\
                       
                     &&  & PB                                  &  0.942(0.190)& 0.920(0.121)& 0.907(0.080)\\
%                       & PBD-2                                & 0.910(0.175)& 0.902(0.116)& 0.904(0.079)\\
\hline

\multirow{18}{*}{1.5}&\multirow{9}{*}{1.5}&\multirow{3}{*}{0.01}& Breslow                                     & 0.955(0.293)& 0.953(0.191)& 0.940(0.130)\\ && & Efron                                        & 0.957(0.294)& 0.954(0.193)& 0.950(0.131)\\
                       
                   &&    & PB                                 & 0.951(0.291)& 0.949(0.192)& 0.945(0.131)\\
%                       & PBD-2                                 & 0.918(0.259)& 0.924(0.174)& 0.925(0.121)\\
\cline{3-7}
&&\multirow{3}{*}{0.1}& Breslow                                     & 0.768(0.249)& 0.546(0.162)& 0.234(0.109)\\  && & Efron                                       & 0.896(0.259)& 0.832(0.170)& 0.708(0.115)\\
                       
                 &&      & PB                                  & 0.947(0.288)& 0.939(0.184)& 0.923(0.122)\\
%                       & PBD-2                                & 0.893(0.245)& 0.892(0.162)& 0.892(0.110)\\
\cline{3-7}
&&\multirow{3}{*}{0.2}& Breslow                                     & 0.418(0.215)& 0.128(0.141)& 0.008(0.095)\\ &&  & Efron                                      & 0.700(0.226)& 0.489(0.148)& 0.232(0.100)\\
                       
                &&       & PB                                  & 0.932(0.276)& 0.909(0.174)& 0.885(0.114)\\
%                       & PBD-2                                & 0.899(0.252)& 0.896(0.166)& 0.885(0.113)\\
\cline{2-7}

& \multirow{9}{*}{2} &\multirow{3}{*}{0.01}& Breslow                                     & 0.943(0.276)& 0.921(0.177)& 0.878(0.119)\\ && & Efron                                        & 0.951(0.278)& 0.943(0.179)& 0.931(0.121)\\
                       
                    &&   & PB                                 & 0.949(0.281)& 0.949(0.183)& 0.944(0.124)\\
%                       & PBD-2                                 & 0.895(0.235)& 0.902(0.155)& 0.900(0.107)\\
\cline{3-7}
&&\multirow{3}{*}{0.1}& Breslow                                     & 0.379(0.199)& 0.099(0.126)& 0.004(0.083)\\ &&  & Efron                                       & 0.585(0.206)& 0.323(0.131)& 0.099(0.086)\\
                       
                     &&  & PB                                  & 0.937(0.273)& 0.909(0.170)& 0.860(0.111)\\
%                       & PBD-2                                & 0.868(0.216)& 0.849(0.140)& 0.804(0.094)\\
\cline{3-7}
&&\multirow{3}{*}{0.2}& Breslow                                     & 0.098(0.163)& 0.004(0.105)& 0.000(0.070)\\  && & Efron                                      & 0.256(0.168)& 0.059(0.108)& 0.002(0.072)\\
                       
                    &&   & PB                                  & 0.910(0.265)& 0.854(0.161)& 0.755(0.102)\\
%                       & PBD-2                                & 0.881(0.236)& 0.844(0.152)& 0.774(0.103)\\
\hline
\end{tabular}
\end{center}
\end{table}

%%%%%%%%%%%%%%%%%%%%%%%%%%%%%%%%%%%%%%%%%%%%%%%%%%%%%%%%%%%%%%%%%%%%%%%%%%%%%%%%%
\section{Real Applications}\label{sec:data_analysis}
%%%%%%%%%%%%%%%%%%%%%%%%%%%%%%%%%%%%%%%%%%%%%%%%%%%%%%%%%%%%%%%%%%%%%%%%%%%%%%%%%

In Section \ref{sec:simul}, we demonstrated that for many cases, $\betahatPB$ surpasses $\betahatB$ and $\betahatE$ in terms of reduced $|\text{Bias}|$ and RMSE, and better confidence interval coverages. This is particularly evident when $\tau$ is large or when there is considerable variation in the $p_{ij}$ values. This finding depicts an important insight: there are positive relationships between $\tau$, ${\sum_{i\in \R(t_{(j)})}p_{ij}^2}/n_j$, and the inaccuracies of $\betahatB$ and $\betahatE$ in estimating $\truebeta$. Here we confirm these relationships using real datasets, anticipating that a large $\tau$ or a high ${\sum_{i\in \R(t_{(j)})}p_{ij}^2}/n_j$ leads to an increase in the RMSE for $\betahatB$ and $\betahatE$, demonstrating the advantage of $\betahatPB$ over these estimators.

For the ease of comparison, we first transformed the variables in all the datasets as follows. All the covariates, except for binary ones, are standardized to have mean 0 and standard deviation 1. The observed times $\{t_i\}_{i=1}^n$ are first scaled to all lie within the interval $[0,1]$. Then to create ties at different levels, we group the scaled times $\{t_i\}_{i=1}^n$ with different choices of $\tau$. Specifically, let $\tau \in \{0, 0.01, 0.02, \dots, 0.25\}$. When $\tau = 0$, the original observed times are used directly, i.e., $t_i^\ast = t_i$. For other values of $\tau$, $t_i^\ast$ is defined as $t_i^\ast = \lceil t_i / \tau \rceil \tau$.

For each $\tau$, we fit $\betahatB$, $\betahatE$, and $\betahatPB$ using the data $\{t_i^\ast, \delta_i, \xvec_i\}_{i=1}^n$. Due to lack of truth in real applications, we consider an estimation accuracy metric defined with respect to $\betahatPB$, considering its theoretical accuracy and consistently strong performance in simulations. In particular, we define the estimation discrepancy (ED) of an estimator $\betahat$ from $\betahatPB$ as
$$\max_{l \in \{1,\cdots,d\}}\left\{\exp\left(|\sbetahat_{l} - \sbetahat_{\text{pb},l}|\right) - 1\right\},$$ where $\betahat$ can be either $\betahatB$ or $\betahatE$. To gain more insights on the estimation accuracy performance, we also record the sum of squared hazards (SSH), defined as $\frac{1}{k}\sum_{j=1}^k\frac{1}{n_j}\sum_{i\in \R(t_{(j)})}\widehat{p}_{ij}^2$, which is the upper bound in Remark \ref{rmk:Lecam} with $\widehat{p}_{ij}$ being the evaluations of $p_{ij}$ at $\betahatPB$ and $\lambdahat_{\text{pb},j}$. Based on our findings in theory and simulation, we expect to observe a positive relationship of ED versus $\tau$ or SSH.

 Besides the measures introduced above, it is also desirable to evaluation the performance of the methods via goodness-of-fit metrics. One such metric is the APL \(L(\cdot, \lambdavec, \zeta)\) with \(\lambdavec\) estimated by $\lambdavechat_{\text{pb}}$. The evaluations of the APL at the corresponding estimates of $\betavec$ yield \(L_{\text{b}} = L(\betahatB, \lambdavechat_{\text{pb}}, \zeta)\), \(L_{\text{e}} = L(\betahatE, \lambdavechat_{\text{pb}}, \zeta)\), and \(L_{\text{pb}} = L(\betahatPB, \lambdavechat_{\text{pb}}, \zeta)\), respectively for the three methods.

\subsection{Male Laryngeal Cancer Patients Study}\label{sec:real_appl_larynx}
The original data are available as larynx dataset from R package ``$\texttt{KMsurv}$". Conducted at a Dutch hospital from 1970 to 1978, this study was reported by \cite{Larynx_dataset}. It includes data on $90$ male patients diagnosed with larynx cancer, documenting the time (in years) to death or censoring after their initial treatment. The dataset also records the covariates such as the age at diagnosis, the year of diagnosis, and the disease stage (four stages in total). More information can be found in \cite{klein03}. For our analysis, we selected age and indicators for Stages 3 and 4 as covariates. 

\begin{figure}%[H]
    % 9 figures
    \centering
    \includegraphics[scale=0.25]{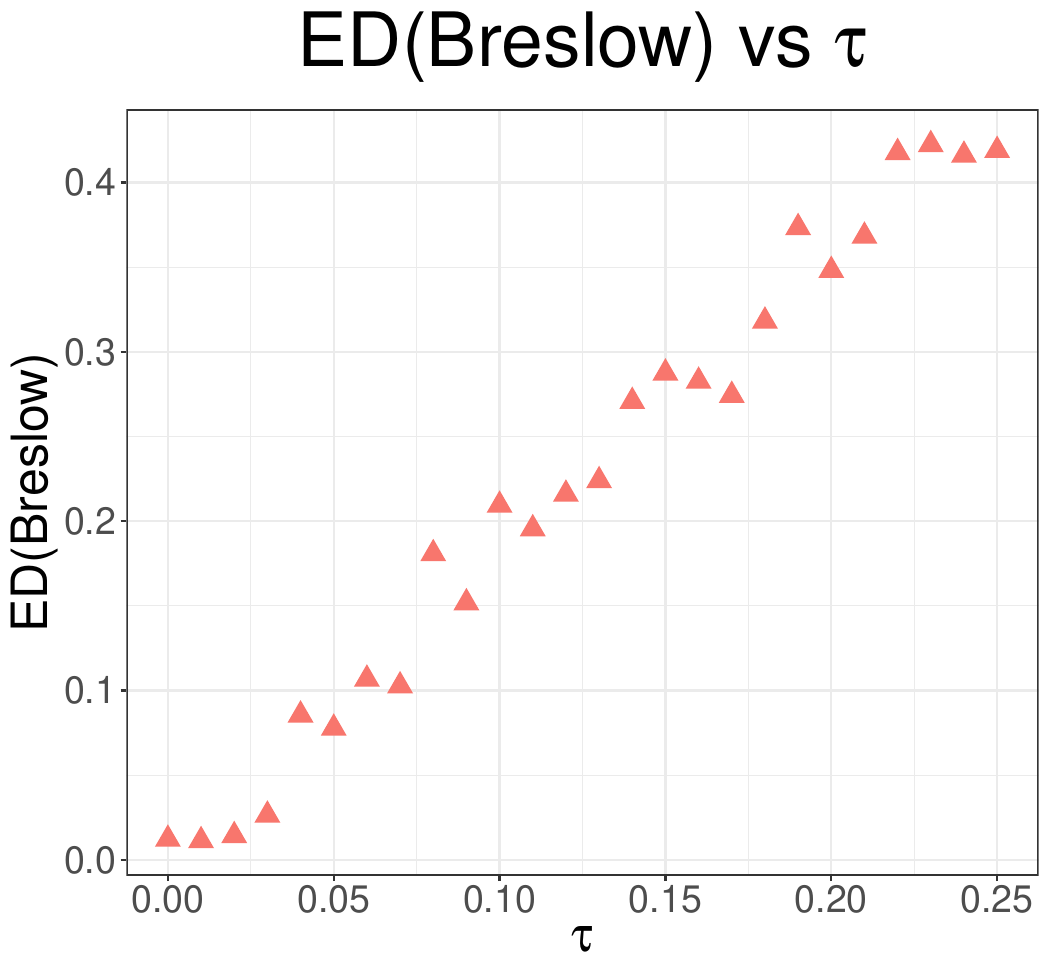}
    \includegraphics[scale=0.25]{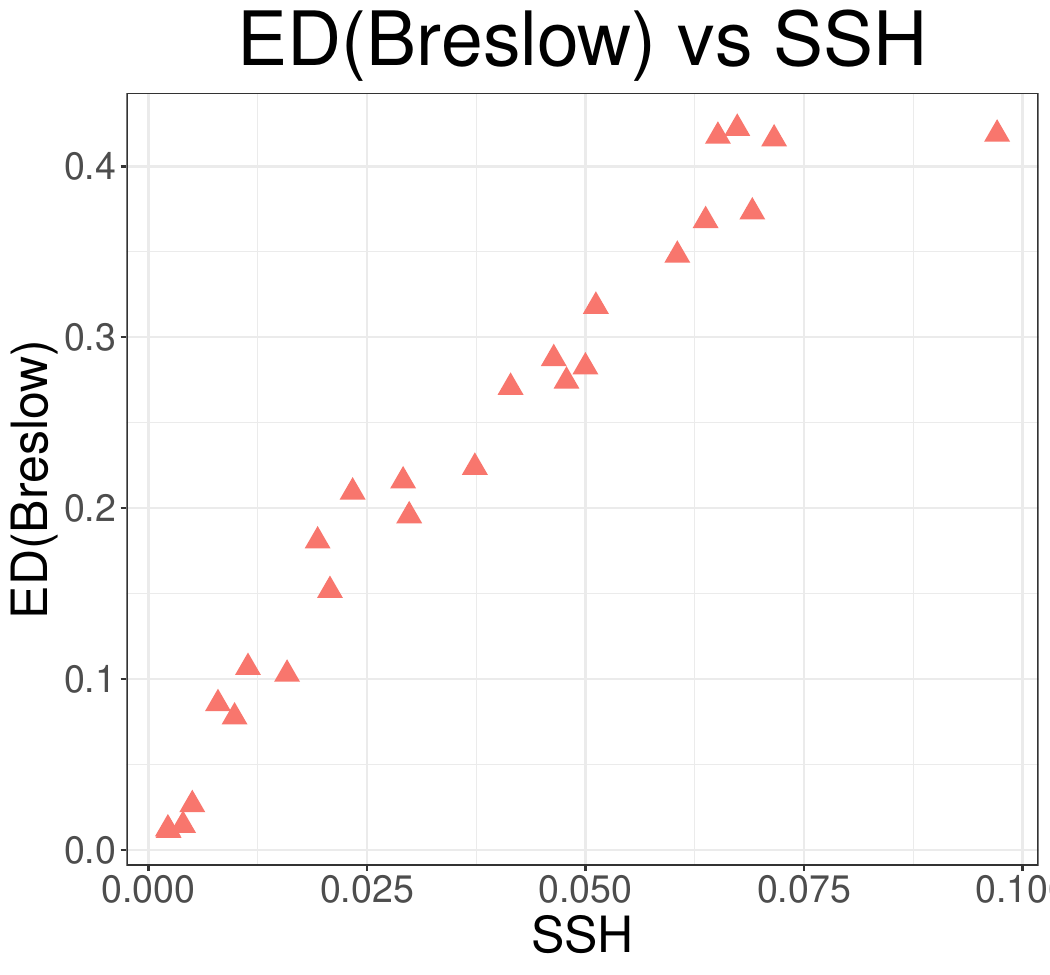}
    \includegraphics[scale=0.25]{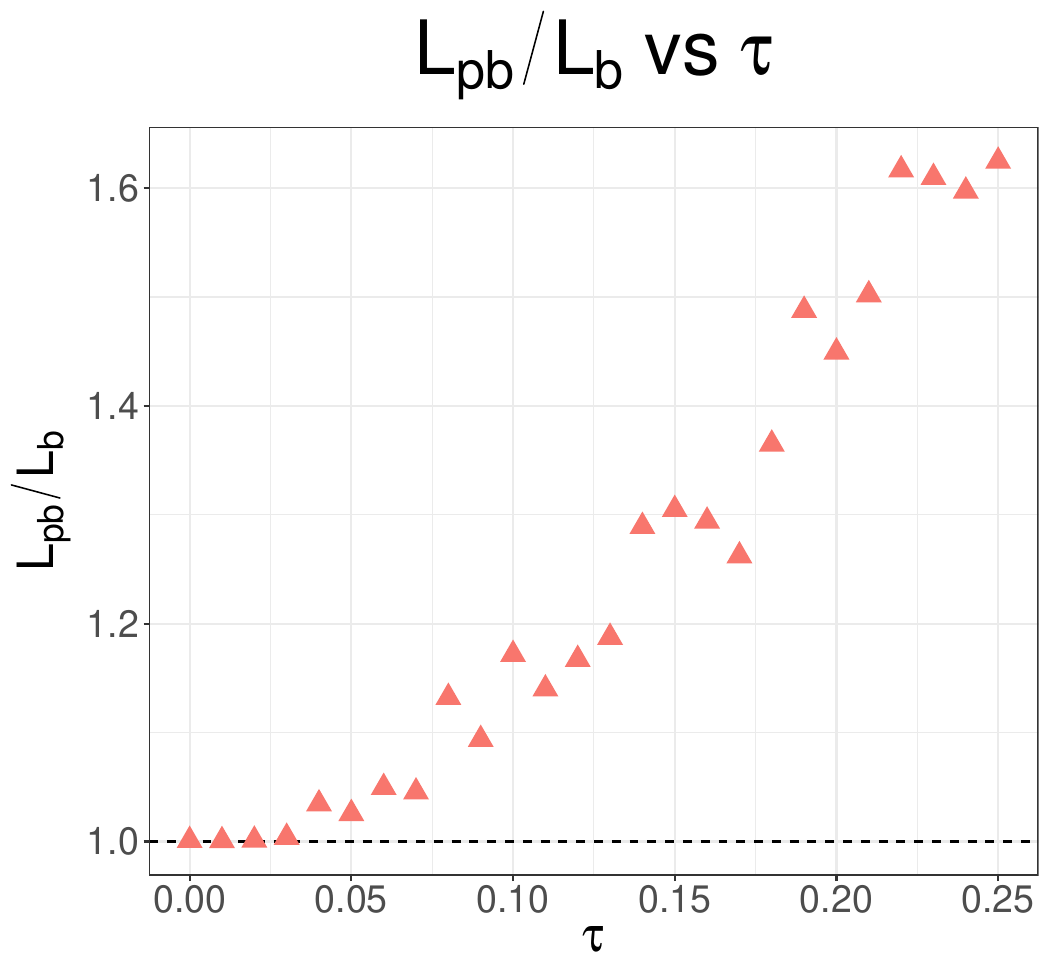}\\
    \includegraphics[scale=0.25]{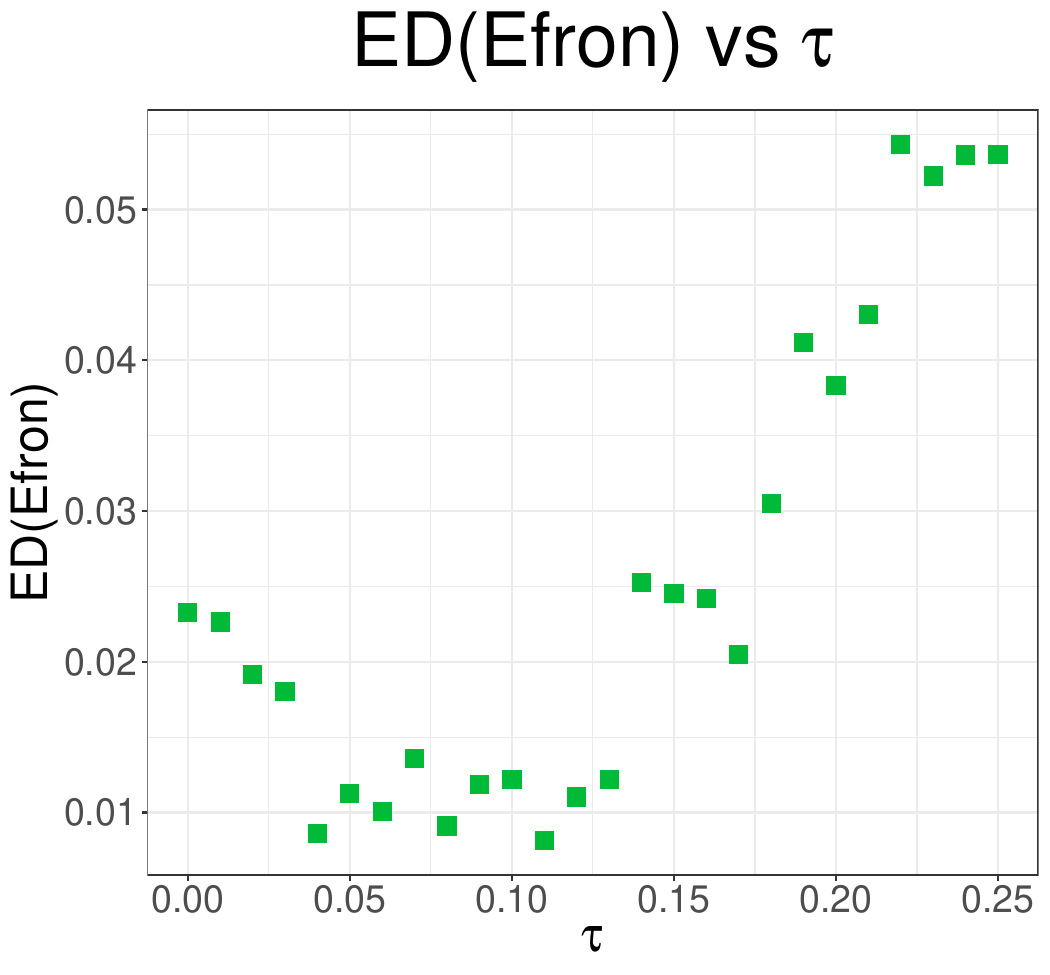}
    \includegraphics[scale=0.25]{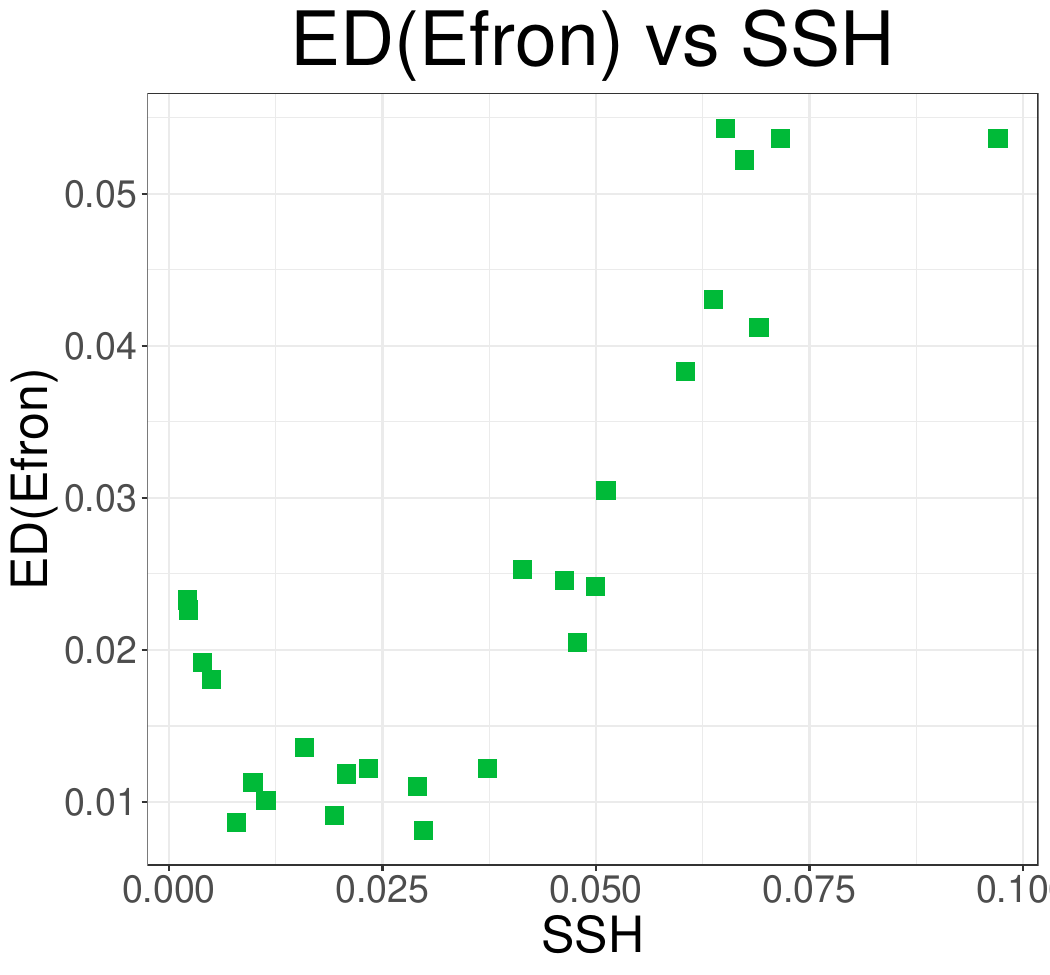}
    \includegraphics[scale=0.25]{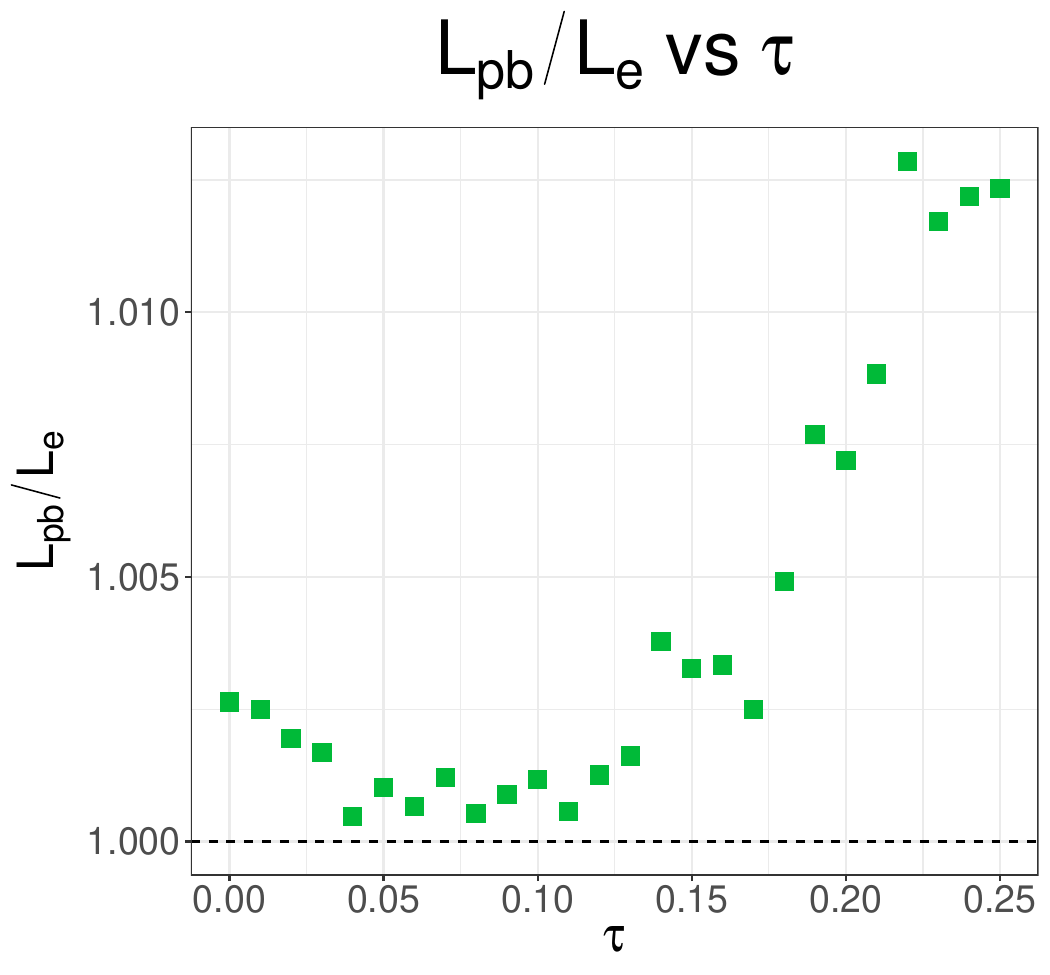}\\
    \includegraphics[scale=0.25]{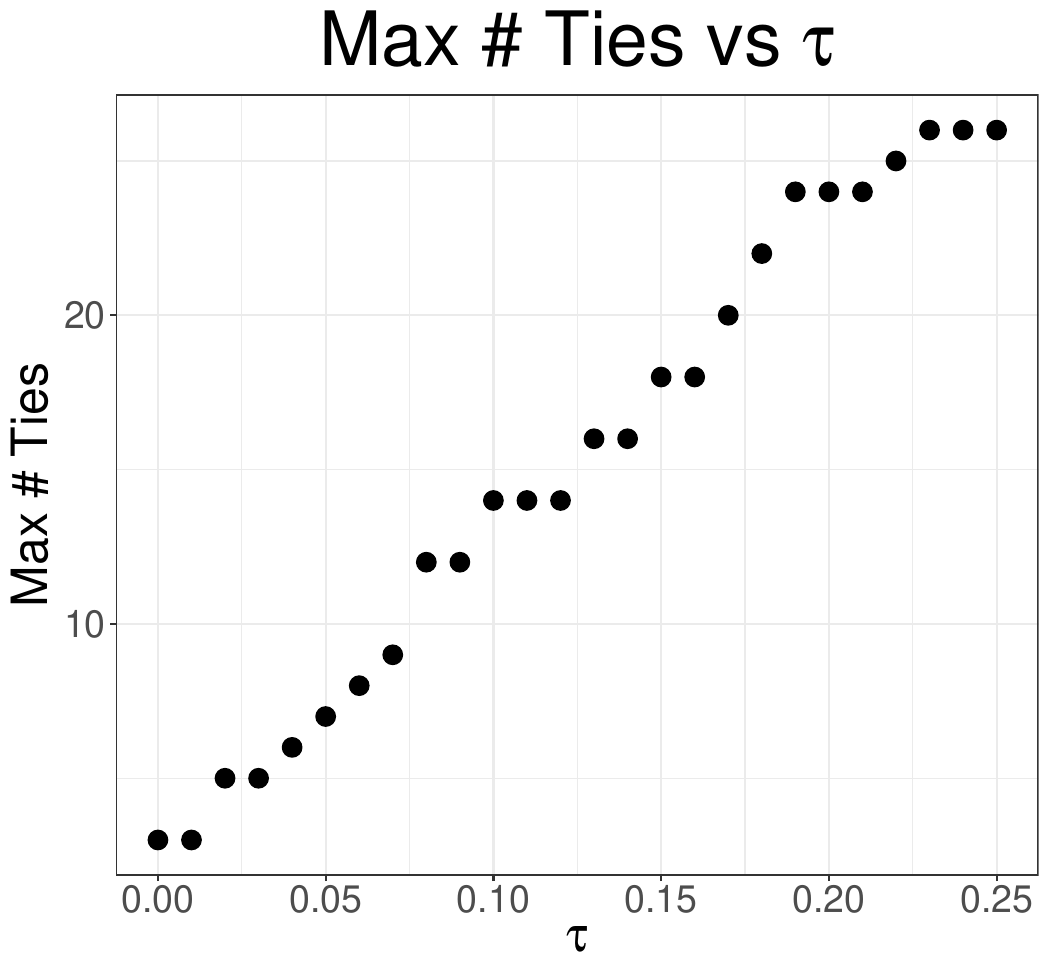}
    \includegraphics[scale=0.25]{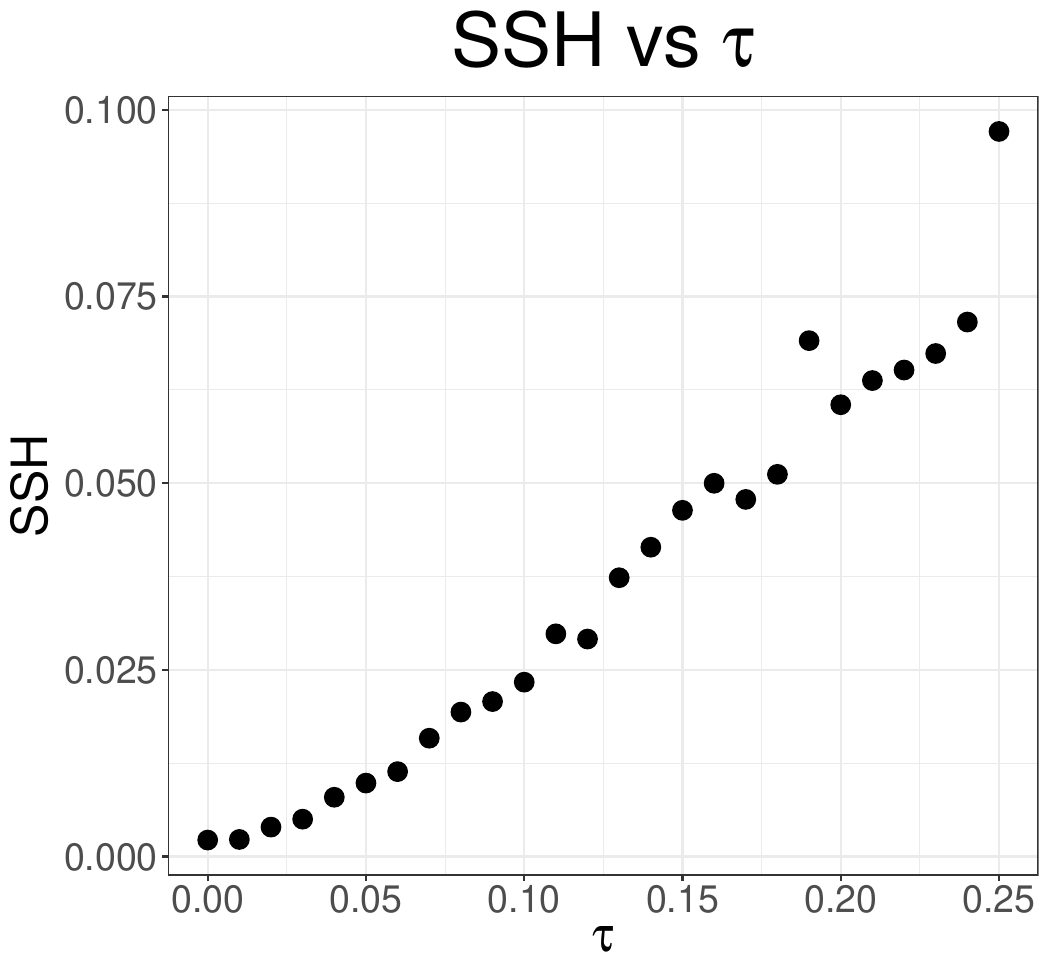}
    \includegraphics[scale=0.25]{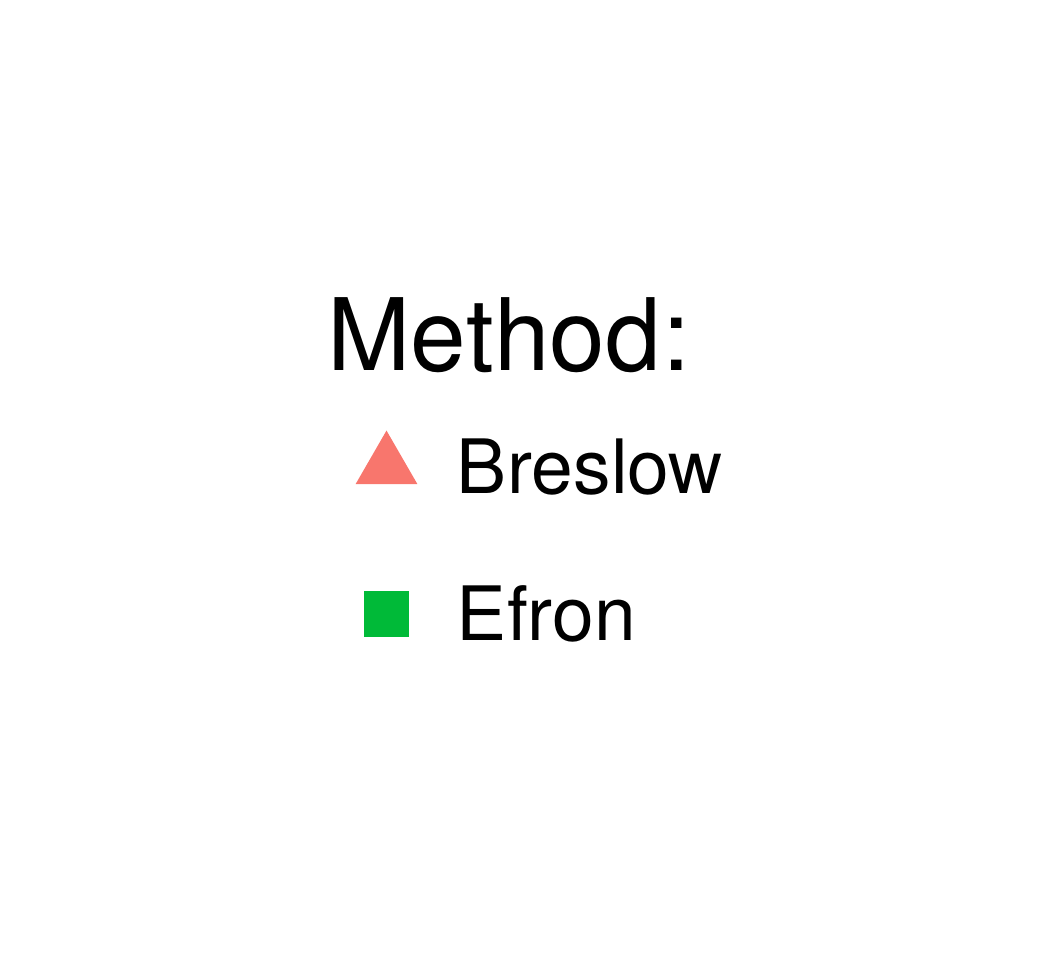}
    \caption{Results for the male laryngeal cancer patients study.}\label{Fig_Larynx}
\end{figure}

As shown in Figure \ref{Fig_Larynx}, the study exhibits the desired positive relationships between the recorded values. In this example, the deviation of the Breslow estimator from the PB distribution estimator increases at an linear rate against $\tau$ or SSH throughout the whole range, while the deviation of the Efron estimator from the PB distribution estimator increases after $\tau\ge 0.125$ or $\text{SSH}\ge 0.375$. The PB distribution estimator shows its advantage against the others in the domain of higher $\tau$ or higher SSH again. 
For the goodness-of-fit, our method delivers slightly higher APLs than the Breslow and Efron methods when $\tau$ is small, and has a clear advantage when $\tau$ increases to higher values.

Supplementary Table S1 indicates that the estimated \(\beta_j\) values are similar across all the methods for small \(\tau\). However, as \(\tau\) increases, Breslow’s estimates diverge significantly from those of our PB distribution based estimates. Although the Efron estimates show closer alignment with our estimates, their discrepancy with our estimates also grows as \(\tau\) increases when viewed alongside ED. On the other hand, the standard error estimates are similar across all the methods, regardless of the \(\tau\) values.

\begin{figure}%[H]
    % 9 figures
    \centering
    \includegraphics[scale=0.25]{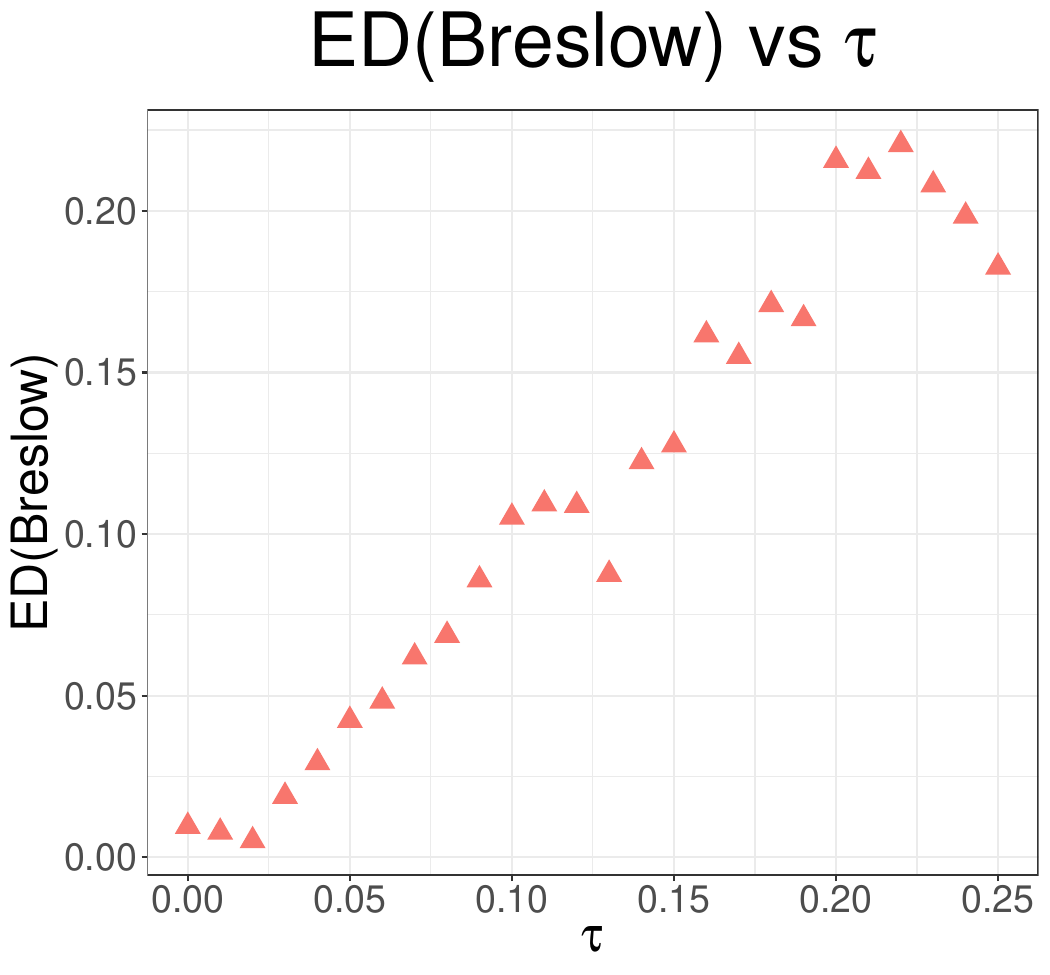}
    \includegraphics[scale=0.25]{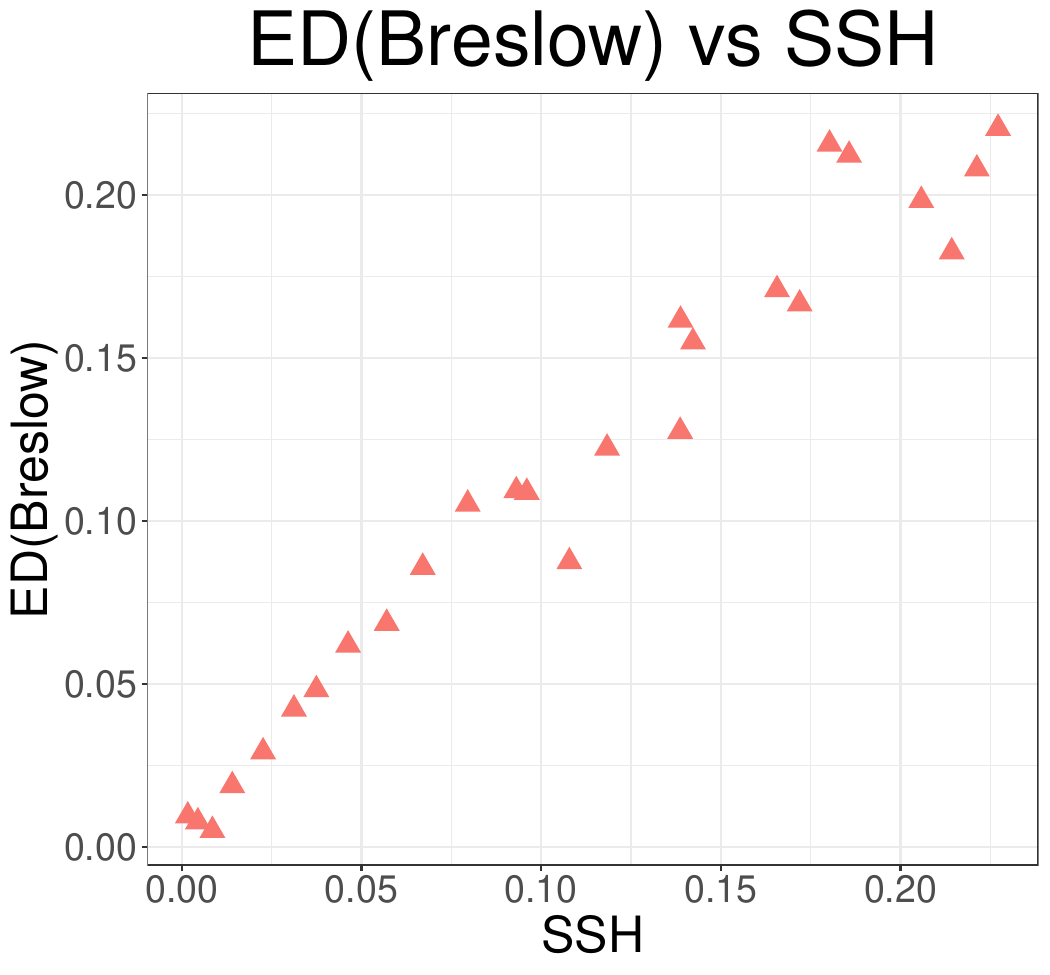}
    \includegraphics[scale=0.25]{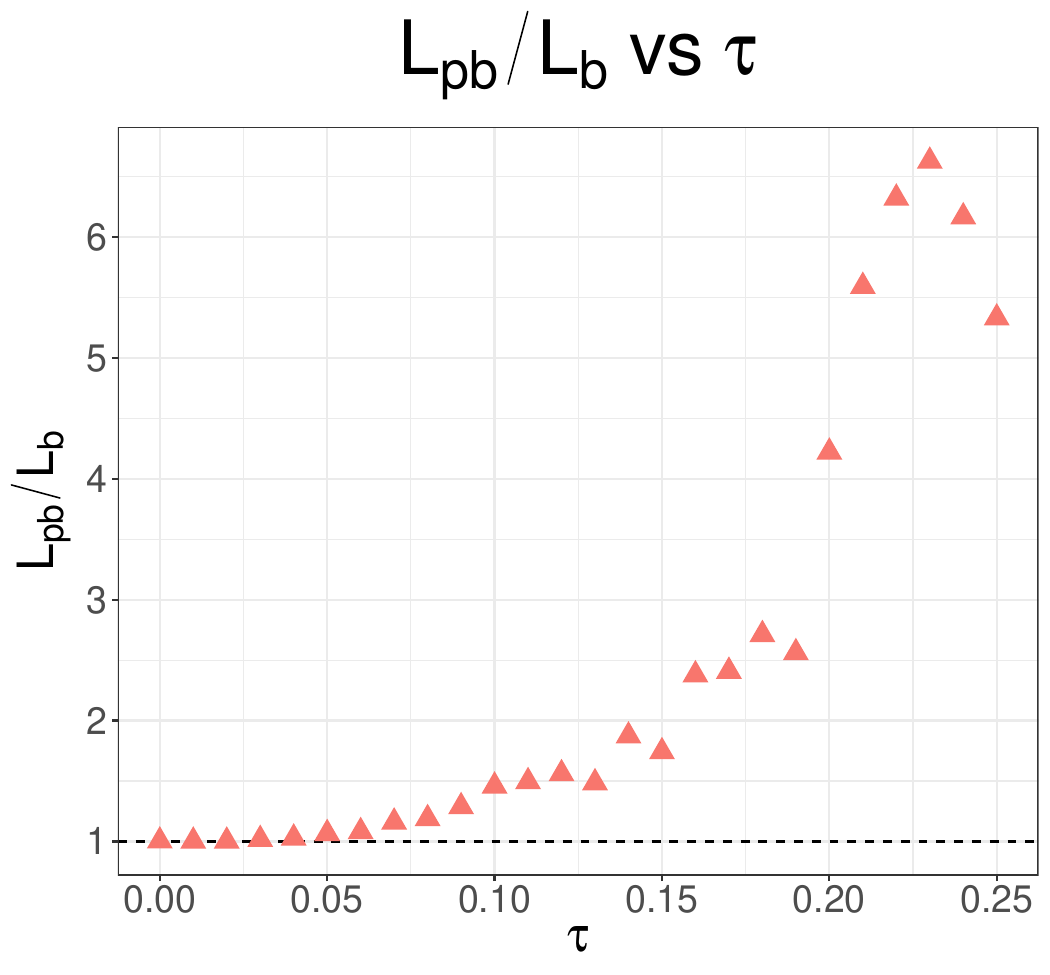}\\
    \includegraphics[scale=0.25]{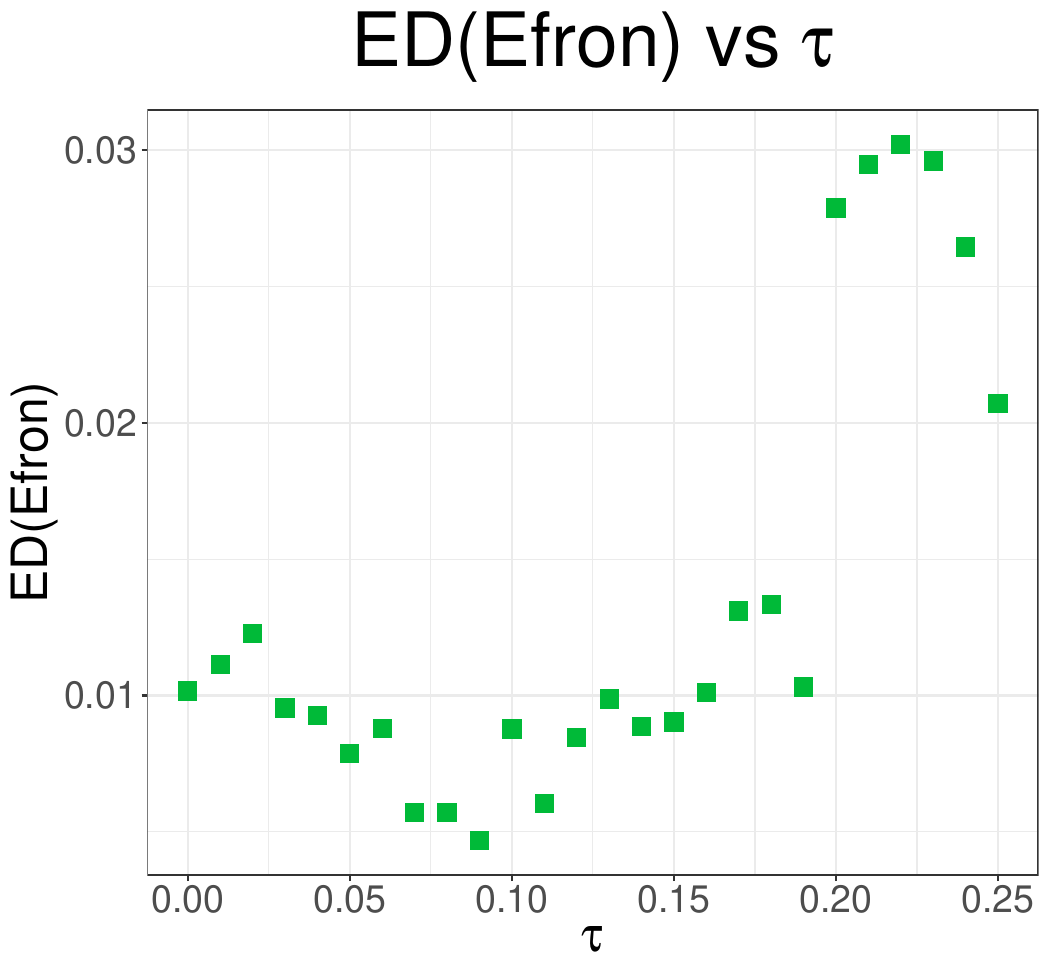}
    \includegraphics[scale=0.25]{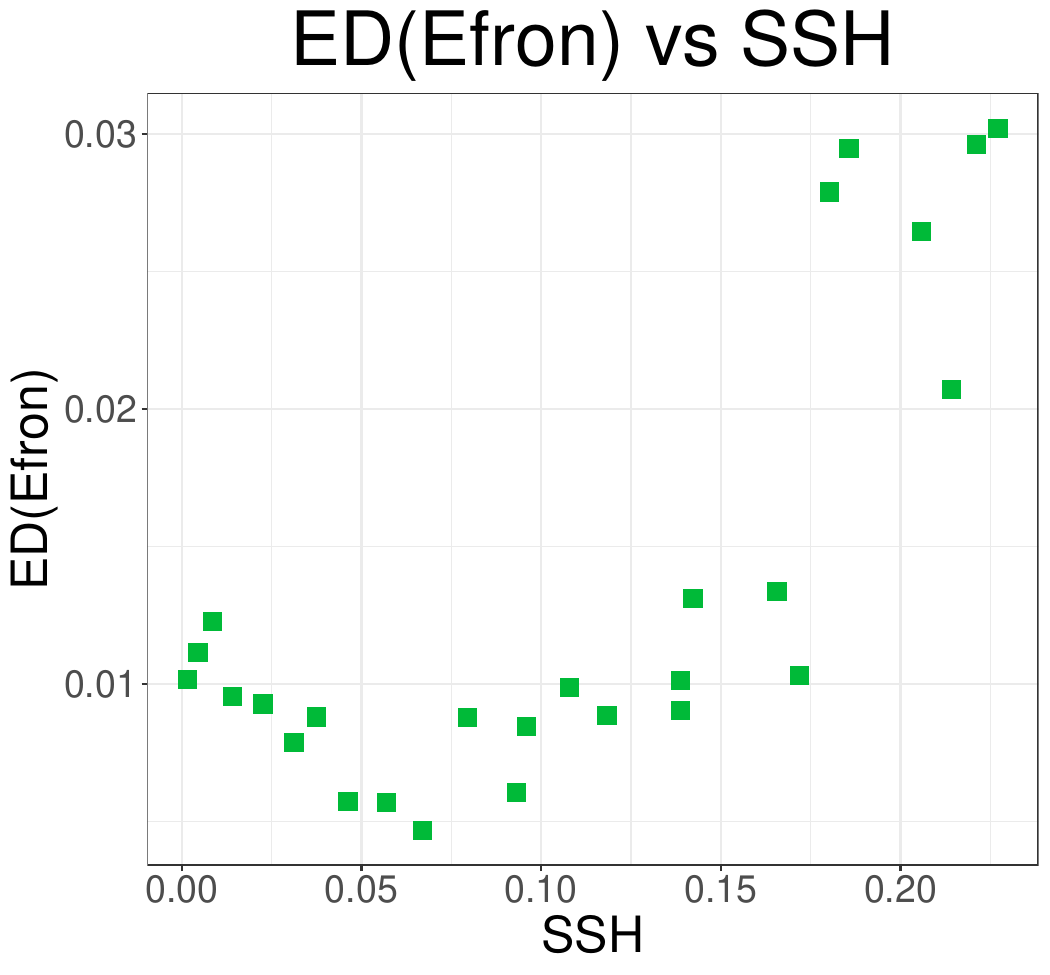}
    \includegraphics[scale=0.25]{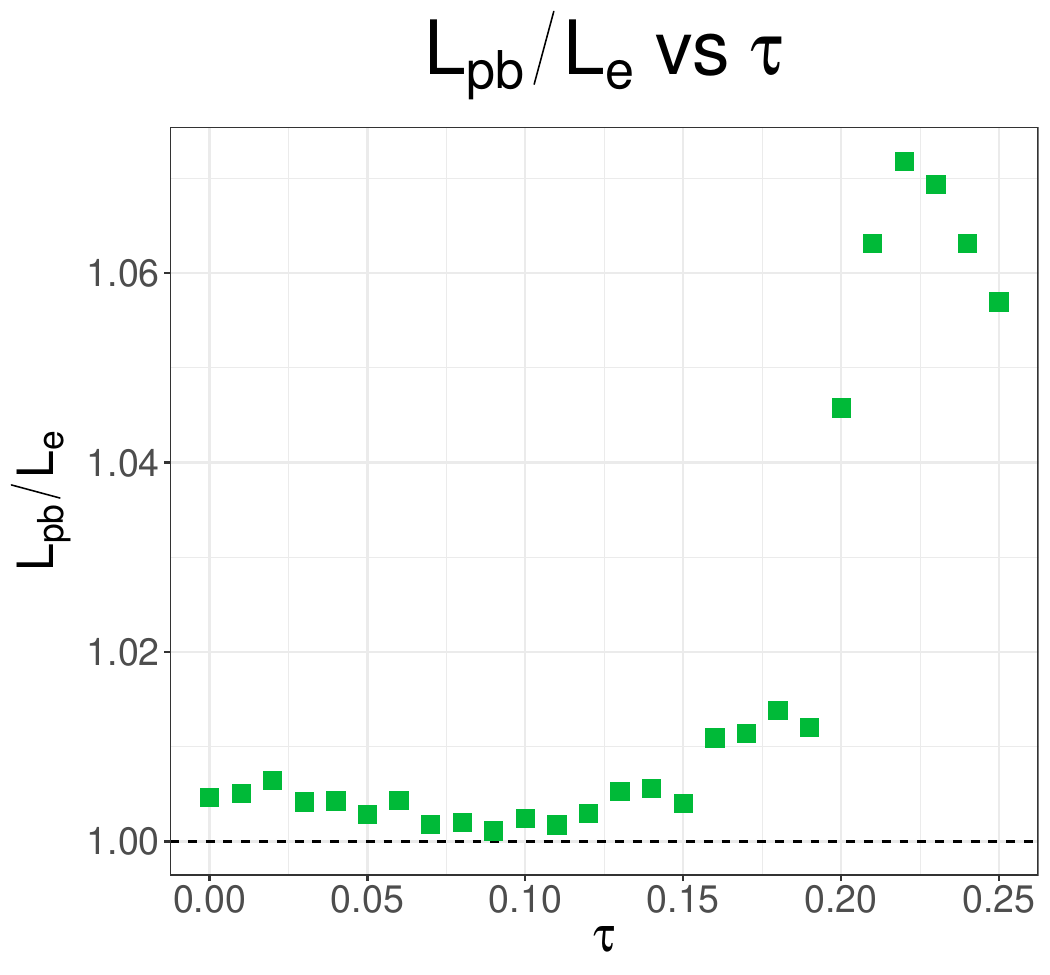}\\
    \includegraphics[scale=0.25]{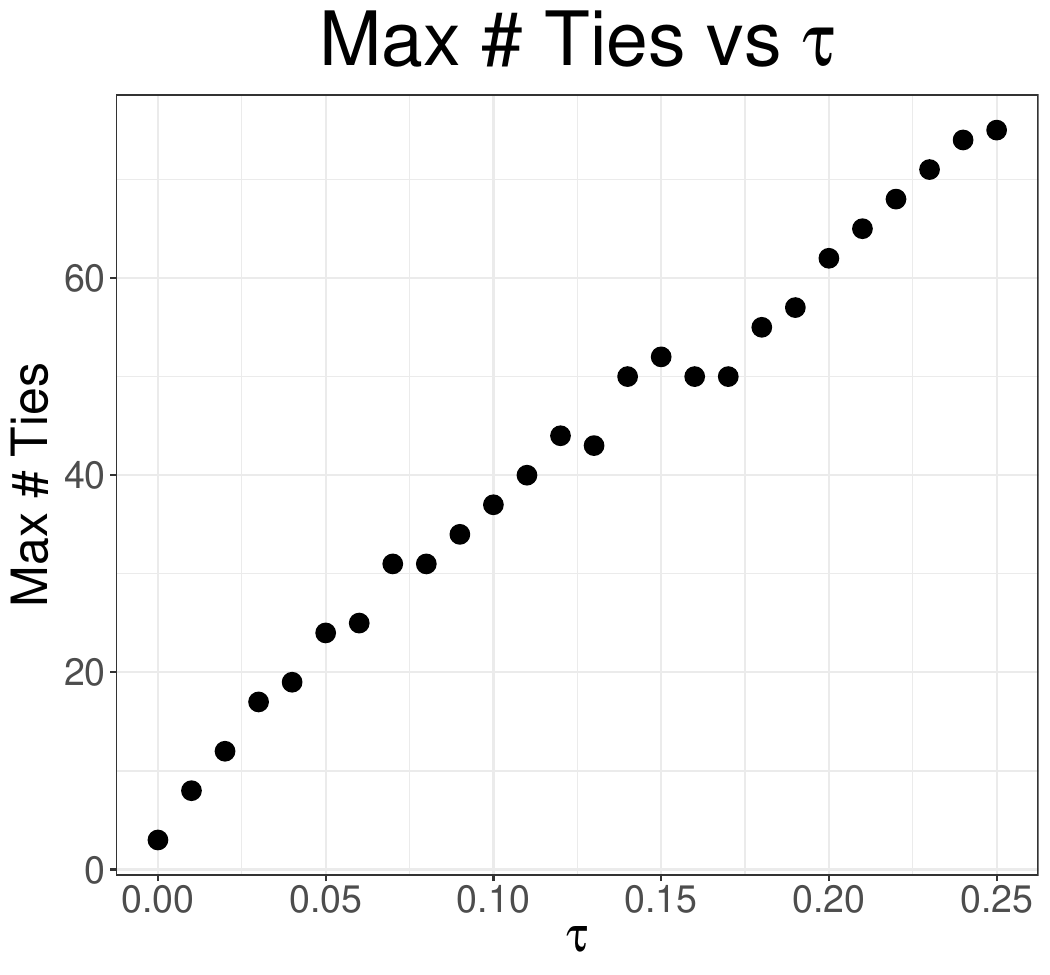}
    \includegraphics[scale=0.25]{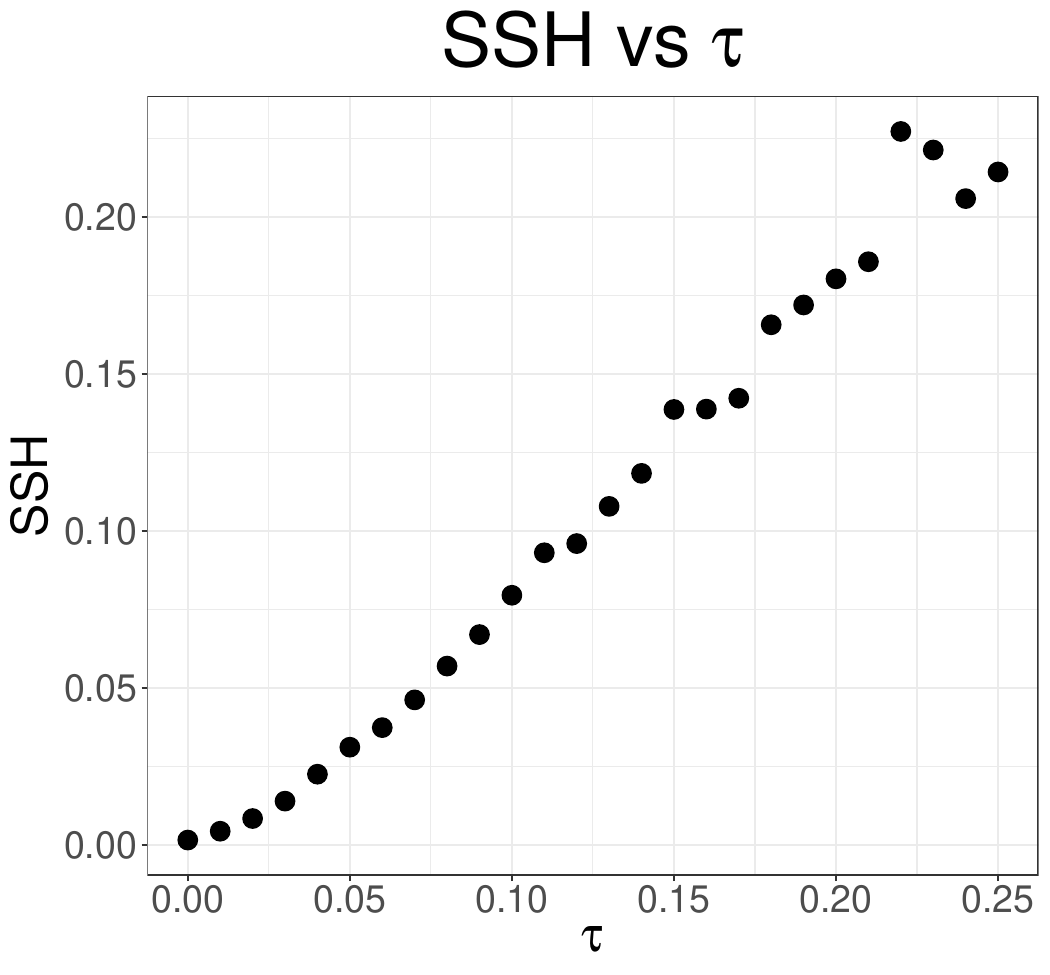}
    \includegraphics[scale=0.25]{{./figures_CSDA/CSDA_legend_real}.pdf}
    \caption{Results for the north central cancer treatment group lung cancer study.}\label{Fig_lung}
\end{figure}

\subsection{North Central Cancer Treatment Group Lung Cancer Study}\label{sec:real_appl_lung}
The lung dataset, also available in the R package ``$\texttt{survival}$", originates from a study conducted by the north central cancer treatment group. This dataset captures survival data for patients with advanced lung cancer, noting the time (in days) to death or censoring for $228$ patients. The covariates we consider here are sex, ECOG performance score as assessed by a physician, Karnofsky performance scores (both patient-rated and physician-rated ones), and weight loss over the last six months (in pounds). For more details, see \cite{Lung_dataset}. 

Figure \ref{Fig_lung} illustrates the desired positive relationships among the recorded values in this study. The trends are similar to what we observe in the laryngeal cancer study except that the extra drastic increases of the Efron estimator's ED happens respectively at $\tau = 0.2$ or $\text{SSH}= 0.175$. The goodness-of-fit comparison and estimation results in Supplementary Table S2 are similar to those in the male laryngeal cancer patients study. However, larger ratios of  the APLs using the PB distribution versus the others are observed for large \(\tau\) values in this study compared to the laryngeal cancer patients study.
\\

%\subsection{General Finding}
{\bf General Finding}: Through the examples with varying tie levels, we have further confirmed our finding in theory and simulation: $\betahatPB$ could estimate $\truebeta$ more accurately than existing methods in datasets with heavy ties or datasets with large variability among covariates. The PB distribution method also outperforms the existing methods in goodness-of-fit when there are heavy ties.

\section{Conclusion}\label{sec:Conclusion}
In this paper, we propose a new approach for accurately computing the partial likelihood for the Cox model, which integrates the original idea of APL and the recent development in efficient calculation of PB distribution. We also thoroughly study the properties of the new computing method. That is, our consistency and asymptotic normality for the new approach can cover not only grouped data with ties, but also continuous data without ties. Our numerical results show that due to reduction in bias, the new approach outperforms existing methods in reducing RMSE and improving coverage of confidence interval for the estimated coefficients, especially when data have many ties or the variation among the risk scores is high. Therefore, we recommend using the new method rather than the existing methods for these cases. One can see that the choice of $\lambdavechat$, which is used as substitution of $\lambdavec$ in APL, is very flexible, only requiring $\lambdavechat$ to satisfy a mild condition in Assumption \ref{assump:efron}.

Without loss of generality and for notation simplicity, we only consider non-time-varying covariates in this paper. The results can be easily extended to time-varying covariates. For our future theory, we might try to relax order condition for $\tau$ needed in the asymptotic theory of $\betahatPB$. Under the grouped continuous failure time model, to have consistency and asymptotic normality, the PB distribution estimator $\betahatPB$ requires $\tau \asymp 1/n$. Meanwhile, the Breslow estimator $\betahatB$ requires $\tau \rightarrow 0$ for consistency and $n^{1/2}\tau \rightarrow 0$ for asymptotic normality, which are much relaxed conditions than $\tau \asymp 1/n$. However, in our numerical example, $\betahatPB$ performs much better than $\betahatB$ under a large $\tau$, i.e., lots of ties. The current theory for $\betahatPB$ only allows a non-diverging number of ties, i.e., $\sup_{t \in \Omega}d(t)=\bigO_{\text{P}}(n\tau)=\bigO_{\text{P}}(n\times1/n)=\bigO_{\text{P}}(1)$, but we might be able to allow a diverging number of ties for $\betahatPB$ in the theory considering its high performances in numerical examples with a large $\tau$ and, correspondingly, lots of ties.

Another direction of future research is to extend this idea to more complex survival models such as the competing risk Cox model, calculating the exact partial likelihood using a Poisson multinomial distribution (PMD). As \citet{lin2022poisson} have proposed an efficient calculation for the PMD, we can use it for calculating the exact partial likelihood. Similar statistical theories in this paper can also be considered.

 As demonstrated in our numerical studies, the PB distribution method not only dominates the existing methods in the case of heavy ties, but also delivers competitive performance in the case of no or fewer ties. In addition, the dominance can also happen when covariate effects or covariate variances are large. However, the exact thresholds of these model parameters for such dominance can be hard to determine and most likely vary in different settings. Therefore, a practical guideline we would like to provide is to always fit a PB distribution method alongside the existing methods whenever there are ties in the data. When discrepancies are observed among the three methods, the PB distribution estimate, whose robustness is demonstrated here, is the one we would recommend.

Here we treat ties as grouped data from underlying continuous times and focus on proportional hazards modeling of the underlying continuous time hazard function. An alternative approach based on such grouped-continuous-times assumption, called discrete-time models that include parametric and machine learning classification methods, is discussed in \cite{Discrete_survival_2022}. However, their approach directly models the discretized hazards instead of the underlying continuous hazard function. The latter form of hazard, as our focus here, is often of more interest in most survival analysis studies.

%  The \backmatter command formats the subsequent headings so that they
%  are in the journal style.  Please keep this command in your document
%  in this position, right after the final section of the main part of
%  the paper and right before the Acknowledgements, Supplementary Materials,
%  and References sections.

%  This section is optional.  Here is where you will want to cite
%  grants, people who helped with the paper, etc.  But keep it short!

%%%%%%%%%%%%%%%%%%%%%%%%%%%%%%%%%%%%%%%%%%%%%%%%%%%%%%%%%%%%%%%%%%%%%%%%%%%%%%%%%%%%%%%%%%%%%%%%%%%%%%%%%%%%%%%%%%%%%%%%%%%%
\section*{Supplementary Materials}

The following supplementary materials are available online.

\begin{description}
\item[Additional details:] Supplementary materials include technical proofs and additional numerical results provided as a PDF file. Additionally, an R package implementing the proposed method is available, along with R code for simulation studies and real applications based on the package.
\end{description}
\par
%%%%%%%%%%%%%%%%%%%%%%%%%%%%%%%%%%%%%%%%%%%%%%%%%%%%%%%%%%%%%%%%%%%%%%%%%%%%%%%%%%%%%%%%%%%%%%%%%%%%%%%%%%%%%%%%%%%%%%%%%%%%
\section*{Acknowledgements}
The authors sincerely thank the Editor, the Associate Editor, and the three anonymous reviewers for their valuable feedback, which has greatly improved the content of this paper. The authors acknowledge the Advanced Research Computing program at Virginia Tech for providing computational resources. 
The second author was supported in part by the COS Dean's Discovery Fund (Award: 452021) and the Data Science Faculty Fellowship (Award: 452118) at Virginia Tech.
\par

%%%%%%%%%%%%%%%%%%%%%%%%%%%%%%%%%%%%%%%%%%%%%%%%%%%%%%%%%%%%%%%%%%%%%%%%%%%%%%%%%%%%%%%%%%

\appendix

%%%%%%%%%%%%%%%%%%%%%%%%%%%%%%%%%%%%%%%%%%%%%%%%%%%%%%%%%%%%%%%%%%%%%%%%%%%%%%%%%
\section{Existing APL Approximation Methods}\label{sec:exisingmethods}
%%%%%%%%%%%%%%%%%%%%%%%%%%%%%%%%%%%%%%%%%%%%%%%%%%%%%%%%%%%%%%%%%%%%%%%%%%%%%%%%%
When there are no ties (i.e., $d_j$=1), the APL is
\begin{align}
L_j(\truebeta, \lambda_j)&=\frac{A_j(\truebeta, \lambda_j)}{B_j(\truebeta, \lambda_j)}=\frac{p_{j_1,j}(\truebeta,\lambda_j) \prod_{i\in \R(t_{(j)})\setminus \{j_1\} }\left(1-p_{ij}(\truebeta,\lambda_j)\right)}{\sum_{i\in\R(t_{(j)})}\left\{p_{ij}\prod_{\ell\in \R(t_{(j)})\setminus \{i\}}(1-{p_{\ell j}})\right\}}.
\end{align}
When $\lambda_j$ is small enough, say $\sup_j\lambda_j=\bigO_{\text{P}}(\tau)$, $p_{ij} \approx \exp(\xvec_i^{\text{T}}\truebeta)\lambda_j$ by the Taylor expansion. Then $\sup_{ij}p_{ij}=\bigO_{\text{P}}(\tau)$, $1-p_{ij}\approx 1$, and \begin{align}\label{eqn:PL.no.ties}
L_j(\truebeta, \lambda_j)\approx\frac{\exp(\xvec_{j_1}^{\text{T}}\truebeta)\lambda_j}{\sum_{i\in\R(t_{(j)})}\exp(\xvec_{i}^{\text{T}}\truebeta)\lambda_j}=
\frac{\exp(\xvec_{j_1}^{\text{T}}\truebeta)}{\sum_{i\in\R(t_{(j)})}\exp(\xvec_{i}^{\text{T}}\truebeta)},
\end{align}
which is the approximate partial likelihood \eqref{eqn:PL.CA.j} used in literature \citep{Cox1972}.

When there are ties, the Cox correction \citep{Cox1972} uses the following partial likelihood term,
\begin{align}\label{eqn:exact_Cox}
L_j(\truebeta, \lambda_j)\approx
\frac{\exp\left(\sum_{i\in\D(t_{(j)})}\xvec_{i}^{\text{T}}\truebeta\right)}{\sum_{\A_{d_j}\in\mathscr{F}_{d_j}}\exp\left(\sum_{i\in \A_{d_j}}\xvec_{i}^{\text{T}}\truebeta\right)}.
\end{align}
We have $d_j$ ties at $t_{(j)}$, and subject $j_1,\dots,j_{d_j}$ failed at $t_{(j)}$. Let $Q_j$ be the set of $d_j!$ permutations of the subjects ${j_1,\dots, j_{d_j}}$. Let $S^j_{l}=\{s^j_{l1},\dots, s^j_{ld_j}\}$ be the $l$th element in $Q_j$. We have $Q_j=\{S^j_{1},\dots,S^j_{d_j!}\}$ and $\R(t_{(j)},S^j_l,m)=\R(t_{(j)})\setminus\{s^j_{l1},\dots,s^j_{l,m-1}\}$. The Kalbfleisch-Prentice correction \citep{KalbfleischPrentice1973} uses the following partial likelihood term,
\begin{align}\label{eqn:exact_KP}
L_j(\truebeta, \lambda_j)\approx\frac{1}{d_j!}
\sum_{l=1}^{d_j!}
\frac{\exp\left(\sum_{i\in\D(t_{(j)})}\xvec_{i}^{\text{T}}\truebeta\right)}{\prod_{m=1}^{d_j}\left\{\sum_{i\in\R(t_{(j)},S^j_l,m)}\exp(\xvec_{i}^{\text{T}}\truebeta)\right\}},
\end{align} which is based on the average partial likelihood contribution at $t_{(j)}$.

The Breslow correction \citep{Breslow1974} uses the following partial likelihood term, \begin{align}\label{eqn:breslow}
L_j(\truebeta, \lambda_j)\propto \frac{\exp\left(\sum_{i\in\D(t_{(j)})}\xvec_{i}^{\text{T}}\truebeta\right)}{\left\{\sum_{i\in\R(t_{(j)})}\exp(\xvec_{i}^{\text{T}}\truebeta)\right\}^{d_j}}.
\end{align} We denote by $\betahatB$ and $\lambdahat_{\text{b}j}={d_j}/\{ \sum_{i\in\R(t_{(j)})}\exp(\xvec_{i}^{\text{T}}\betahatB)\}$ the Breslow estimates for $\truebeta$ and $\lambda_j$, respectively, where $\widehat{\Lambda}_{\text{b}}(t) = \sum_{j=1}^k \lambdahat_{\text{b}j} \mathds{1}(t_{(j)} \leq t)$.

The Efron correction \citep{Efron1977} uses the following partial likelihood term, \begin{align}\label{eqn:efron}
L_j(\truebeta, \lambda_j)\propto \frac{\exp\left(\sum_{i\in\D(t_{(j)})}\xvec_{i}^{\text{T}}\truebeta\right)}
{\prod_{\ell=0}^{d_j-1}\left\{\sum_{i\in\R(t_{(j)})}\exp(\xvec_{i}^{\text{T}}\truebeta)-\ell \overline{A}\left( \truebeta, t_{(j)} \right) \right\}},
\end{align}
where $\overline{A}(\truebeta, t_{(j)})=\sum_{i\in\D(t_{(j)})}\exp\left(\xvec_{i}^{\text{T}}\truebeta\right)/d_j$. We denote the Efron estimates for $\truebeta$ and $\lambda_j$ by $\betahatE$ and $\lambdahat_{\text{e}j}$, respectively. In particular,
$\lambdahat_{\text{e}j}=\sum_{\gamma=0}^{d_j-1}\big\{ \sum_{i\in\R(t_{(j)})}$ $\exp(\xvec_{i}^{\text{T}}\betahatE) -\gamma \overline{A}(\betahatE, t_{(j)})\big\}^{-1}$,
 where $\widehat{\Lambda}_{\text{e}}(t) = \sum_{j=1}^k \lambdahat_{\text{e}j} \mathds{1}(t_{(j)} \leq t)$. The Cox correction, the Kalbfleisch-Prentice correction, the Breslow correction, and the Efron correction are all the same as the approximate partial likelihood \eqref{eqn:PL.CA.j} when $d_j=1$.
%\textbf{YH: Youngjin, double check this formula. I feel it is \eqref{eqn:efron_baseline} is different from in R package. }

%\textbf{Youngjin}: give formula for $\lambdahat_{\text{e}j}$.

%\textbf{Youngjin}: please finish these two.

%The Efron correction is...

%The Exact correction is...

\section{Asymptotic Notations}\label{Sec:asymp_notation}
Here we introduce some asymptotic notations. Without loss of generality, \( C \) or \( \C \), including versions with subscripts, represent constants in $(0,\infty)$. To avoid confusion with censoring values denoted by \( C_i \), we refrain from using numerical subscripts for \( C \) when representing constants. We denote by $\|A\|_{\infty}$  the infinite norm of a scalar, vector, or matrix. Let $X_n$ and $Y_n$ be random scalar, vector, or matrix. We write $X_n=\bigO_{\text{P}}(Y_n)$ if $\pr(\|X_n\|_\infty > C \|Y_n\|_\infty) \rightarrow 0$ as $n \rightarrow \infty$ for some $C>0$ and $X_n=o_{\text{P}}(Y_n)$ when $\pr(\|X_n\|_\infty > \epsilon \|Y_n\|_\infty) \rightarrow 0$ as $n \rightarrow \infty$ for all $\epsilon >0$. One can see that $\|\cdot\|_\infty$ in definitions of $\bigO_{\text{P}}(\cdot)$ or $o_{\text{P}}(\cdot)$ can be substituted by $\|\cdot\|_2$ for vectors with finite, fixed, and therefore non-diverging dimensions as $n \rightarrow \infty$, i.e., $\|a\|_\infty \leq \|a\|_2 \leq \sqrt{m} \|a\|_\infty$ for $a \in \mathbb{R}^m$. Let $\dgoto$, $\Pgoto$, and $\asgoto$ stand for convergence in distribution, convergence in probability, and almost sure convergence, respectively. Note that summation over empty set is set to be $0$ and product over empty set is set to be $1$.

\section{Assumptions}\label{sec:assumptions}

Recall that $\zeta$ is the ending time of the failure time study, which we assume, for the sake of notational simplicity, to be a multiple of the grouping parameter $\tau$.

\begin{assumption}\label{assump:enough_sample_cmg}
Recall that $n(t)$ and $d(t)$ are respectively the numbers of at-risk subjects and failed subjects at time $t$. We assume $n(\zeta)-d(\zeta) \geq 1$.
\end{assumption}
Note that Assumption \ref{assump:enough_sample_cmg} is basically saying that there is at least one censored subjects in the end of the study, which holds for most datasets. This assumption basically allows $n(\zeta)$ or $n(\zeta)-d(\zeta)$ appearing as the denominators of fractions.
\begin{assumption}\label{assump:conti_no_out_lier}
The covariates $\xvec_1,\dots,\xvec_n$ are {\color{black}deterministic and are} elements of some compact set $\mathcal{M}$.
\end{assumption}

Assumption \ref{assump:conti_no_out_lier} implies $\sup_{i}\|\xvec_i\|_{\infty} \leq \rho$ for some $\rho>0$, or $\sup_i\|\xvec_i\|_\infty=\bigO(1)$, that is, there are no extreme outliers in the model. This is commonly assumed in the survival analysis literature.

\begin{assumption}\label{assump:Lipshitz}
We assume $\sup_{t\in\Omega}d \Lambda_0^*(t) = \bigO(\tau)$, where $\Lambda_0^*(\cdot)$ is the discretized baseline cumulative hazard function defined in \eqref{eqn:discretized_cox_base_chf}.
\end{assumption}
One can see that Assumption \ref{assump:Lipshitz} can be generally satisfied for common cumulative hazard functions. One example is the cumulative hazard function of the Weibull distribution $\left(\Lambda_0(t)=gt^m\right)$ with the parameters $g>0$ and $m \geq 1$. To check this, one can see that $d\Lambda^\ast_0(t)/\tau$ is non-decreasing function on $\Omega$ and $$\frac{d\Lambda^\ast_0(\zeta)}{\tau}=\frac{ g \zeta^m - g(\zeta-\tau)^m }{\tau}=\frac{  g(\zeta-\tau)^m -g \zeta^m  }{-\tau} \rightarrow \frac{\partial}{\partial\zeta} g\zeta^m =gm\zeta^{m-1}$$ as $\tau \rightarrow 0$ as $n \rightarrow \infty$. So, $0 \leq d\Lambda^\ast_0(\tau)/\tau \leq d\Lambda^\ast_0(2\tau)/\tau \leq \cdots \leq d\Lambda^\ast_0(\zeta)/\tau = \{\sup_{t \in \Omega} d \Lambda_0^\ast(t)\}/\tau= \bigO(1)$, directly implying $ \sup_{t \in \Omega}d\Lambda^\ast_0(t)=\bigO(\tau)$. To see point-wise boundedness for some individual $t \in \Omega$, as $d\Lambda^\ast_0(\zeta)/\tau \rightarrow gm\zeta^{m-1}$ as $\tau \rightarrow 0$ by $n \rightarrow \infty$, $d\Lambda^\ast_0(\zeta)\asymp \tau$, which is also $\bigO(\tau)$, and $d\Lambda^\ast_0(\tau)=g\tau^m$ is $o(\tau)$ for $m>1$ and $\asymp \tau$ for m=1, which are also $\bigO(\tau)$.

\begin{assumption}\label{assump:order_nt}
We assume $\sup_{t \in [0,\zeta]}|n(t)/n - y(t)| \asgoto 0$ as $n \rightarrow \infty$ with $y(t) \in (0,1]$ for all $t \in [0,\zeta]$.
\end{assumption}
Assumption \ref{assump:order_nt} means that there are always a nonzero fraction of subjects at risk even throughout the study. This is a reasonable assumption since most survival data contain a good portion of censored subjects at the end of the study.

\begin{assumption}\label{assump:tie_order}
We assume $\sup_{t\in\Omega}dN(t) =\bigO_{\text{P}} \left( \sup_{t \in \Omega}\left\{ \E(dN(t)|\mathcal{F}_{t^-}) \right\} \right).$
\end{assumption}
Assumption \ref{assump:tie_order}, together with Assumptions \ref{assump:enough_sample_cmg} -- \ref{assump:order_nt}, indicates that $$\sup_{t \in \Omega} \left\{\E(dN(t)|\mathcal{F}_{t^-}) \right\}=\sup_{t \in \Omega} dA(t)=\sup_{t \in \Omega} \left\{\sum_{i=1}^n Y_i(t)d\Lambda^{\ast}(t;\xvec_i) \right\} = \bigO_{\text{P}}(n\tau),$$ implying that $\sup_{t\in\Omega}d(t)=\sup_{t\in\Omega}dN(t)=\bigO_{\text{P}}(n\tau)$. Considering the fact that the number of elements in $\Omega$ is $\zeta/\tau$ and the number of total events is $\bigO_{\text{P}}(n)$, it is natural that $\sup_{t\in\Omega}dN(t)=\bigO_{\text{P}}(n/(\zeta/\tau))=\bigO_{\text{P}}(n\tau)$, which means the number of ties are balanced for all $t \in \Omega$, i.e., no extreme number of ties for some $t \in \Omega$.

For martingales $G(\cdot)$ and $H(\cdot)$, we define predictable variation process as $\langle G \rangle(t)=\int_0^t d\langle G\rangle(s)$, where $d\langle G\rangle(t)=\Var(d G(t)|\mathcal{F}_{t^-})$ and the predictable covariation process as $\langle G,H \rangle(t)=\int_0^t d\langle G,H\rangle(s)$, where $d\langle G,H\rangle(t)=\textrm{Cov}(d G(t),$ $ d H(t)|\mathcal{F}_{t^-})$.

\begin{assumption}\label{assump:grp_indep}
We assume
$$\E(dN_i(t)dN_l(t)|\mathcal{F}_{t^-})=\E(dN_i(t)|\mathcal{F}_{t^-})\E(dN_l(t)|\mathcal{F}_{t^-}),$$ when $i \neq l$ for all $t \in [0, \zeta]$.
\end{assumption}
Assumption \ref{assump:grp_indep} is common in the survival analysis literature. One can see that it implies $d\langle M_i, M_l\rangle(t)=0$, when $i \neq l$ for all $t \in [0, \zeta]$.

Define the processes 
\begin{align}
 &S^{(0)}(\truebeta,t)=\sum_{i=1}^n Y_i(t)\exp(\xvec_i^{\text{T}} \truebeta),\\
 &S^{(1)}(\truebeta,t) =\frac{\partial}{\partial\truebeta}S^{(0)}(\truebeta,t) =\sum_{i=1}^n Y_i(t)\xvec_i \exp(\xvec_i^{\text{T}} \truebeta),\text{ and}\\
 &S^{(2)}(\truebeta,t) = \frac{\partial^2}{\partial\truebeta \partial\truebeta^{\text{T}}}S^{(0)}(\truebeta,t) =\sum_{i=1}^n Y_i(t)\xvec_i\xvec_i^{\text{T}} \exp(\xvec_i^{\text{T}} \truebeta).
\end{align}

\begin{assumption}\label{assump:grp_consistency}
There exists an open convex neighborhood $\mathcal B$ of $\truebeta$ and functions {\color{black}$s^{(j)}(\cdot,\cdot)$}, $j=0,1,2$ defined on $\mathcal B \times [0,\zeta]$, that satisfy the following four conditions.
\begin{enumerate}
    \item {\color{black}$\int_0^\zeta d\Lambda_0(t) < \infty$.}
    \item $\sup_{\tilde{\truebeta} \in \mathcal{B}, t \in [0,\zeta]} \| S^{(j)}(\tilde{\truebeta},t)/n-s^{(j)}(\tilde{\truebeta},t) \|_{\infty} \Pgoto 0$ as $n \rightarrow \infty$ for $j=0,1,2$,
    \item {\color{black}$s^{(0)}(\tilde{\truebeta},t)>m_s>0$} for all $\tilde{\truebeta} \in \mathcal{B}$, $t \in [0,\zeta]$  {\color{black} and $\sup_{\tilde{\truebeta} \in \mathcal{B}, t \in [0,\zeta]}\|s^{(j)}(\tilde{\truebeta},t)\|_\infty$ $ \leq C_S$ for $j=0,1,2$ for some $C_S>0$},
    \item For $j=0,1,2$, $s^{(j)}(\tilde{\truebeta},t)$ are continuous functions of {\color{black}$\tilde{\truebeta} \in \mathcal{B}$ uniformly in} $t \in [0,\zeta]$, and $s^{(1)}(\tilde{\truebeta},t)={\partial}s^{(0)}(\tilde{\truebeta},t)/{(\partial \tilde{\truebeta})}$,  $s^{(2)}(\tilde{\truebeta},t)={\partial^2}s^{(0)}(\tilde{\truebeta},t)/$ ${(\partial \tilde{\truebeta}\partial \tilde{\truebeta}^{\text{T}})}$ {\color{black}for all $\tilde\truebeta\in\mathcal{B}, t\in[0,\zeta]$}, and
    \item $\Sigma(\tilde\truebeta, \zeta)=\int_{0}^{\zeta}v(\tilde{\truebeta},t)s^{(0)}(\tilde{\truebeta},t) d\Lambda_0(t)$ is positive definite for all $\tilde{\truebeta} \in \mathcal{B}$, where $e(\tilde{\truebeta},t)=s^{(1)}(\tilde{\truebeta},t)/s^{(0)}(\tilde{\truebeta},t)$ and $v(\tilde{\truebeta},t)=s^{(2)}(\tilde{\truebeta},t)/s^{(0)}(\tilde{\truebeta},t)-e(\tilde{\truebeta},t)e(\tilde{\truebeta},t)^{\text{T}}$ {\color{black}for all $\tilde\truebeta\in\mathcal{B}, t\in[0,\zeta]$.}
\end{enumerate}
    \end{assumption}
Assumption \ref{assump:grp_consistency} is common in the survival analysis literature. See, for example, Section 8 of \citet{FlemingHarrington1991}. Without loss of generality, there exists $C_{\mathcal{B}}>0$ such that $\|\tilde{\truebeta}\|_2 \leq C_{\mathcal{B}}$ for all $\tilde{\truebeta} \in \mathcal{B}$.

\begin{assumption}\label{assump:order_tau_n}
Let $\tau \asymp 1/n$, which means $\tau=\bigO(1/n)$ and $1/n=\bigO(\tau)$.
\end{assumption}
Assumption \ref{assump:order_tau_n}, together with Assumptions \ref{assump:enough_sample_cmg} -- \ref{assump:tie_order}, indicates $\sup_{t \in \Omega} d(t)=\bigO_{\text{P}}(1)$, which means the supremum number of ties does not diverge with $n$.

\begin{assumption}\label{assump:efron}
For all $t \in \Omega$, $d\widehat{\Lambda}(t) = d(t)/n^\ast(t)$ for some non-negative and non-increasing function $n^\ast(t)$ over $t \in [0,\zeta]$, where $\sup_{t \in \Omega} n^\ast(t)=\bigO_{\text{P}}(n)$ and $\sup_{t \in \Omega} \{1/n^\ast(t)\}=\bigO_{\text{P}}(1/n)$.
\end{assumption} Under Assumptions \ref{assump:enough_sample_cmg} -- \ref{assump:tie_order} and \ref{assump:efron}, one can show that $\sup_{t\in\Omega} d\widehat{\Lambda}(t) = \bigO_{\text{P}}(\tau)$. One can see that $d\Lambdahat_{\text{b}}(\cdot)$ and $d\Lambdahat_{\text{na}}(\cdot)$ satisfy Assumption \ref{assump:efron}, where $\Lambdahat_{\text{na}}(t)=\sum_{j=1}^k \lambdahat_{\text{na},j} \mathds{1}(t_{(j)} \leq t)$. See Section S1.8. Although there is no rigorous proof that $d\Lambdahat_{\text{e}}(\cdot)$ satisfies Assumption \ref{assump:efron}, as the Efron correction is a modification of the Breslow correction and performance of $\lambdavechat_{\text{e}}$ in Supplementary Figure S2 is better than the other baseline hazard function estimators in reducing bias of $\betahatPB$, we use $d\Lambdahat_{\text{e}}(\cdot)$ as $d\Lambdahat(\cdot)$ in our numerical examples in Sections \ref{sec:simul} and \ref{sec:data_analysis}.
{\color{black}
\begin{assumption}\label{assump:grp_pbd_an}
    $X_{\text{pb}}(\cdot,\lambdavechat,\zeta)$ and $X(\cdot,\zeta)$ are strictly concave on $\mathcal{B}$ and there exist $a \in [0,1/2)$, $C_{\text{pb}}>0$, and $C_{\text{b}}>0$ satisfying the following as $n \rightarrow \infty$:
    \begin{enumerate}
        \item $\pr\left(\|\betahatPB-\betahatB\|_2 \leq C_{\text{pb}}n^a\left\{ X_{\text{pb}}(\betahatPB,\lambdavechat,\zeta)- X_{\text{pb}}(\betahatB,\lambdavechat,\zeta) \right\}  \right) \rightarrow 1$.
        \item $\pr\left(\|\betahatB-\betahatPB\|_2 \leq C_{\text{b}}n^a\left\{ X(\betahatB,\zeta)- X(\betahatPB,\zeta) \right\}  \right) \rightarrow 1$.
    \end{enumerate}
\end{assumption}}
One can see that for any $\tilde\truebeta \in \mathcal B$, we have $X_{\text{pb}}(\betahatPB,\lambdavechat,\zeta)- X_{\text{pb}}(\tilde\truebeta,\lambdavechat,\zeta) \geq 0$ and $X(\betahatB,\zeta)- X(\tilde\truebeta,\zeta) \geq 0$ by the definition of $\betahatPB$ and $\betahatB$.
Due to strict concavity, $X_{\text{pb}}(\betahatPB,\lambdavechat,\zeta)- X_{\text{pb}}(\tilde\truebeta,\lambdavechat,\zeta)=0$ only when $\tilde{\truebeta}=\betahatPB$, which also makes $\|\betahatPB-\tilde\truebeta\|_2=0$. Similarly, $X(\betahatB,\zeta)- X(\tilde\truebeta,\zeta)=0$ only when $\tilde{\truebeta}=\betahatB$, which also makes $\|\betahatB-\tilde\truebeta\|_2=0$. {\color{black}Assumption \ref{assump:grp_pbd_an} essentially requires that the difference between $\betahatPB$ and $\betahatB$ is bounded by both the difference between their values in $n^aX_{\text{pb}}(\cdot,\lambdavechat,\zeta)$ and the difference between their values in $n^aX(\cdot,\zeta)$. Considering the fact that $\betahatPB, \betahatB \in \mathcal{B}$, we have $\|\betahatPB-\betahatB\|_2=\bigO_{\text{P}}(1)$. Also, one can show that $n^a\{X_{\text{pb}}(\betahatPB,\lambdavechat,\zeta)- X_{\text{pb}}(\betahatB,\lambdavechat,\zeta)\}=\bigO_{\text{P}}(n^a)$ and $n^a\{X(\betahatB,\zeta)- X(\betahatPB,\zeta)\}=\bigO_{\text{P}}(n^a)$. So this assumption is expected to be satisfied in many scenarios.}

For $j=0,1,2$, define the processes $\mathcal{S}^{(j)}(\truebeta,t)$ similar to $S^{(j)}(\truebeta,t)$ with the at-risk process $Y_i(t)$ replaced by $\mathcal{Y}_i(t)=\mathds{1}(T_i \geq t)$ for the continuous model, where $T_i=\min(\tilde{T}_i,C_i)$.

\begin{assumption}\label{assump:conti_consistency}
Let $\sum_{i=1}^n\mathcal{Y}_i(\zeta)-1 \geq 1$ and  $\sup_{t \in [0,\zeta]}|\sum_{i=1}^n\mathcal{Y}_i(t)/n - \mathscr{Y}(t)| \asgoto 0$ as $n \rightarrow \infty$ with $\mathscr{Y}(t) \in (0,1]$ for all $t \in [0,\zeta]$. There exists an open convex neighborhood $\mathcal B$ of $\truebeta$ and functions $\mathscr{S}^{(j)}(\cdot,\cdot)$, $j=0,1,2$ defined on $\mathcal B \times [0,\zeta]$, that satisfy the following four conditions.
\begin{enumerate}
    \item {\color{black}$\int_0^\zeta d\Lambda_0(t) < \infty$.}
    \item $\sup_{\tilde{\truebeta} \in \mathcal{B}, t \in [0,\zeta]} \| \mathcal{S}^{(j)}(\tilde{\truebeta},t)/n-\mathscr{S}^{(j)}(\tilde{\truebeta},t) \|_{\infty} \Pgoto 0$ as $n \rightarrow \infty$ for $j=0,1,2$,
    \item {\color{black}$\mathscr{S}^{(0)}(\tilde{\truebeta},t)>m_\mathscr{S}>0$} for all $\tilde{\truebeta} \in \mathcal{B}$, $t \in [0,\zeta]$  {\color{black} and $\sup_{\tilde{\truebeta} \in \mathcal{B}, t \in [0,\zeta]}\|\mathscr{S}^{(j)}(\tilde{\truebeta},t)\|_\infty $ $\leq C_\mathscr{S}$ for $j=0,1,2$ for some $C_\mathscr{S}>0$},
    \item For $j=0,1,2$, $\mathscr{S}^{(j)}(\tilde{\truebeta},t)$ are continuous functions of {\color{black}$\tilde{\truebeta} \in \mathcal{B}$ uniformly in} $t \in [0,\zeta]$, and $\mathscr{S}^{(1)}(\tilde{\truebeta},t)={\partial}\mathscr{S}^{(0)}(\tilde{\truebeta},t)/{(\partial \tilde{\truebeta})}$,  $\mathscr{S}^{(2)}(\tilde{\truebeta},t)={\partial^2}\mathscr{S}^{(0)}(\tilde{\truebeta},t)/$ ${(\partial \tilde{\truebeta}\partial \tilde{\truebeta}^{\text{T}})}$  {\color{black}for all $\tilde\truebeta\in\mathcal{B}, t\in[0,\zeta]$}, and
    \item $\Sigma_{\text{c}}(\tilde\truebeta, \zeta)=\int_{0}^{\zeta} v_{\text{c}}(\tilde{\truebeta},t)\mathscr{S}^{(0)}(\tilde{\truebeta},t) d\Lambda_0(t)$ is positive definite for all $\tilde{\truebeta} \in \mathcal{B}$, where $e_{\text{c}}(\tilde{\truebeta},t)=\mathscr{S}^{(1)}(\tilde{\truebeta},t)/\mathscr{S}^{(0)}(\tilde{\truebeta},t)$ and $v_{\text{c}}(\tilde{\truebeta},t)=\mathscr{S}^{(2)}(\tilde{\truebeta},t)/\mathscr{S}^{(0)}(\tilde{\truebeta},t)-e_{\text{c}}(\tilde{\truebeta},t)e_{\text{c}}(\tilde{\truebeta},t)^{\text{T}}$ {\color{black}for all $\tilde\truebeta\in\mathcal{B}, t\in[0,\zeta]$}.
\end{enumerate}
\end{assumption}
As there is at most a single tie in continuous model, Assumption \ref{assump:conti_consistency} is saying that at least one subject is censored at the end of the study.
{\color{black}
\begin{assumption}\label{assump:conti_pbd_an}
    $\chi_{\text{pb}}(\cdot,\lambdavechat)$ and $\chi(\cdot)$ are strictly concave on $\mathcal{B}$ and there exist $\alpha \in [0,1/2)$, $\mathcal{C}_{\text{pb}}>0$, and $\mathcal{C}_{\text{c}}>0$ satisfying the following as $n \rightarrow \infty$:
    \begin{enumerate}
        \item $\pr\left(\|\betahatPB-\widehat{\truebeta}_{\text{c}}\|_2 \leq \mathcal{C}_{\text{pb}}n^{\alpha}\left\{ \chi_{\text{pb}}(\betahatPB,\lambdavechat)-\chi_{\text{pb}}(\widehat{\truebeta}_{\text{c}},\lambdavechat) \right\}  \right) \rightarrow 1$.
        \item $\pr\left(\|\widehat{\truebeta}_{\text{c}}-\betahatPB\|_2 \leq \mathcal{C}_{\text{c}}n^{\alpha}\left\{ \chi(\widehat{\truebeta}_{\text{c}})- \chi(\betahatPB) \right\}  \right) \rightarrow 1$.
    \end{enumerate}
\end{assumption}}

\bibliographystyle{elsarticle-harv} 
\bibliography{ref}

\end{document}